\documentclass[reprint, 
 amsmath,amssymb,
 aps, 
 prx,
 longbibliography,
floatfix,
]{revtex4-2}
\usepackage{graphicx}
\usepackage{dcolumn}
\usepackage{bm}

\usepackage[utf8]{inputenc}
\usepackage[T1]{fontenc}
\usepackage{etoolbox}
\usepackage{xcolor}

\usepackage{amsmath}

\DeclareMathOperator*{\argmax}{argmax}

\begin{document}

\title[]{Optimal sensing on an asymmetric exceptional surface}
\author{Robert L. Cook, Liwen Ko,  K. Birgitta Whaley}
\email{whaley@berkeley.edu}
  \affiliation{Department of Chemistry, University of California, Berkeley, CA 94720, USA}

\date{\today}

\begin{abstract}
We study the connection between exceptional points (EPs) and optimal parameter estimation, in a simple system consisting of two counter-propagating traveling wave modes in a microring resonator.   The unknown parameter to be estimated is the strength of a perturbing cross-coupling between the two modes.  Partially reflecting the output of one mode into the other creates a non-Hermitian Hamiltonian which exhibits a family of EPs, creating an exceptional surface (ES).  We use a fully quantum treatment of field inputs and noise sources to obtain a quantitative bound on the estimation error by calculating the quantum Fisher information (QFI) in the output fields, whose inverse gives the Cram\'er-Rao lower bound on the mean-squared-error of any unbiased estimator.  We determine the bounds for two input states, namely, a semiclassical coherent state and a highly nonclassical NOON state.  We find that the QFI is enhanced in the presence of an EP for both of these input states and that both states can saturate the Cram\'er-Rao bound. We then identify idealized yet experimentally feasible measurements that achieve the minimum bound for these two input states. We also investigate how the QFI changes for parameter values that do not lie on the ES, finding that these can have a larger QFI, suggesting alternative routes to optimize the parameter estimation for this problem.
\end{abstract}

\maketitle

\section{\label{sec:Introduction}Introduction}

The utility of exceptional points (EP) for enhanced sensing has been a matter of debate in recent years~\cite{wiersig_enhancing_sensitivity_frequency_2014, wiersig_sensors_operating_exceptional_2016}.  When a non-Hermitian system with an exceptional point of order $m$ is subject to a perturbation of magnitude $\epsilon$, the splitting of the complex energy eigenvalues scales as $\sqrt[m]{\epsilon}$.~\cite{wiersig_review_2020} This contrasts to an equivalent system with only a diabolic point (DP), that is, when the eigenvalues are degenerate but the eigenvectors remain distinct.  For such a system the eigenvalue splitting only scales linearly with $\epsilon$.  The ratio of the two splittings, $\sqrt[m]{\epsilon}/\epsilon = \epsilon^{1/m - 1}$ diverges in the limit as $\epsilon \rightarrow 0$.  This suggests the possibility that measuring the energy splitting near an EP may give a substantial improvement in detecting small values of $\epsilon$.  However, despite this promise, several theoretical analyses~\cite{lau_fundamental_2018, chen_sensitivity_2019, bao_fundamental_2021, duggan_limitations_2022, loughlin_exceptional-point_2024}, 
have concluded that there is no dramatic increase in sensitivity of an EP based sensor.

In this paper, we address this question by quantifying the sensitivity of a two-mode passive system using the quantum Fisher information (QFI).  We show that there is indeed an enhanced sensitivity relative to the equivalent diabolic system when an EP is present. This enhancement is found to be present in a fully quantum analysis of the system, which includes the inherent noise introduced to the system from the input fields.  

\begin{figure}[htbp]
    \includegraphics[width=1.0\columnwidth]{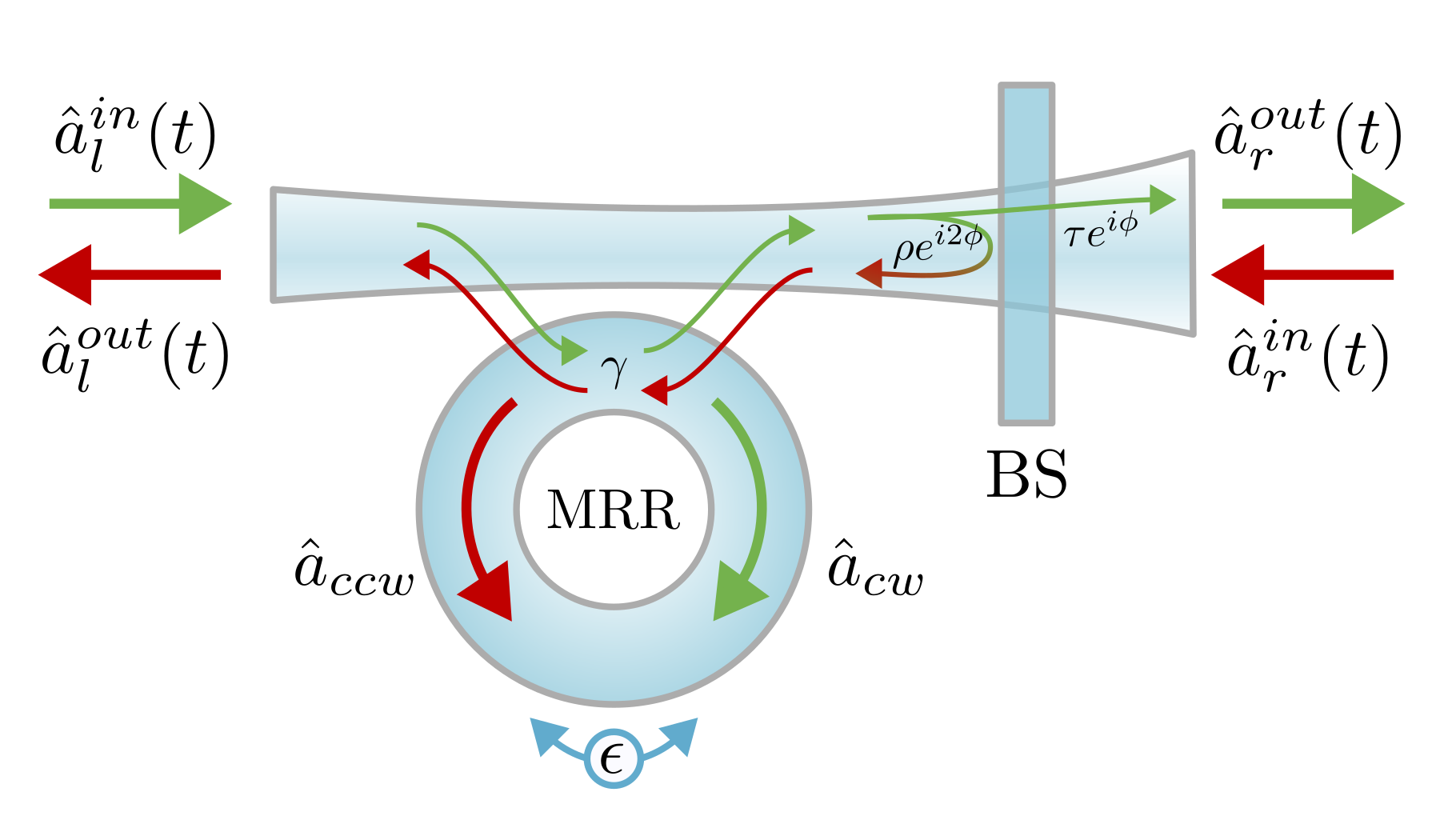}
    \caption{\label{fig:schematic}[color online] Set up for asymmetric coupling in a microring resonator (MRR).  The resonator acts as a cavity that supports clockwise (CW) and counter-clockwise (CCW) whispering gallery modes with quantized field operators $\hat{a}_{cw}$ and $\hat{a}_{ccw}$, respectively.  Left/right input photons (with field operator $\hat{a}_l$/$\hat{a}_r$) propagate in a waveguide and resonantly couple to the cavity modes with the rate $\gamma$. The presence of an external scatterer perturbs the resonator and induces a symmetric cross-coupling between the two whispering gallery modes, parameterized by the real value $\epsilon$.  To the right of the resonator a beam splitter (BS) introduces the asymmetric coupling between the CW and CCW modes by reflecting CW output back as a CCW input. The BS is parameterized by reflection (transmission) coefficient $\rho$ ($\tau$), and phase angle $\phi$.}
\end{figure}

Our system centers around a microring resonator supporting two counter-propagating modes, denoted as clockwise (CW) and counterclockwise (CCW) modes, where each mode couples to a left or right propagating 1D input field. The setup, inspired by the systems studied in refs.~\cite{zhong_sensing_exceptional_surfaces_2019, soleymani_chiral_degenerate_perfect_2022}, is characterized by a continuous two-dimensional surface of exceptional points, i.e., an exceptional surface (ES), and is illustrated in Figure~\ref{fig:schematic}.  The ES in this system arises by partially reflecting the output of one cavity mode into the other. 
This cascading has a nontrivial effect on the system, leading to both coherent and dissipative modifications to the system's evolution.  In fact, it generates a non-Hermitian Hamiltonian with degenerate eigenvalues and eigenvectors that exhibits an EP for a non-zero reflectivity $\rho > 0$.  The reference DP is therefore the $\rho = 0$ configuration, i.e., the case of complete transmission.

The quantity that we wish to estimate in this setup, $\epsilon$, is the strength of a cross-coupling between the two modes, e.g., as might be induced by presence of an impurity embedded in the cavity~\cite{soleymani_chiral_degenerate_perfect_2022, lee_chiral_exceptional_point_2023}.  The presence of such an impurity splits the degeneracy that generates the EP.  Detecting the presence of such an impurity can have significant real-world applications. Previous studies have shown that scattering-induced mode splitting in ultrahigh quality resonators is a useful method for detecting a variety of nanoparticles, ranging from polystyrene spheres~\cite{zhu_onchip_single_nanoparticle_2010}, to single protein complexes~\cite{arnold_shift_whisperinggallery_modes_2003} and individual Influenza A viruses~\cite{vollmer_single_virus_detection_2008}. 

We quantify the sensing enhancement by calculating the quantum Fisher information $\mathcal{I_Q}(\epsilon)$ for all estimates of the parameter $\epsilon$.  The QFI is a useful metric for parameter estimation because the quantum Cram\'er-Rao bound (QCRB) says that the inverse of the QFI gives a lower bound on the mean-squared error (MSE) for any unbiased estimator~\cite{braunstein_statistical_1994}, i.e., 
\begin{equation}
    \text{MSE}_\epsilon \ge \frac{1}{ m\,\mathcal{I_Q}(\epsilon)},
\end{equation}
where $m$ is the number of independent experimental runs used in computing the estimate (see Section \ref{sec:fisher} below).  Since the QFI is intrinsically dependent on the quantum state of the system, we perform the calculation for two different input states, namely a semiclassical coherent state with amplitude $\boldsymbol{\beta}$ and a nonclassical NOON state.  

We show in this work that for a coherent state input, the QFI satisfies the bound $\mathcal{I_Q}(0) \le 64 (1 + \rho)^2 |\boldsymbol{\beta}|^2/\gamma^2$, where $\gamma$ is the system coupling rate and $|\boldsymbol{\beta}|^2$ is the average input photon number.  This shows that for constant photon number, varying the reflectivity between the extremes of $\rho = 0$ and $\rho = 1$, i.e., going from the DP to the EP with a maximal reflection, results in a change in the QFI by at most a factor of 4.  While this finite enhancement is not the unbounded improvement suggested by the eigenvalue scaling, it nevertheless demonstrates a sensing enhancement when operating at an EP rather than the corresponding DP. We show in Appendix~\ref{app:Lau_clerk} that this result is also consistent with a general bound on the sensing signal-to-noise ratio for a perturbed linear network and a coherent state input that was derived in Ref.~\onlinecite{lau_fundamental_2018}.

An intuitive explanation for the existence of enhancement in this situation is the simple fact that the reflection creates a standing wave where the interference between the right and left traveling waves can enhance the total cavity coupling~\cite{dorner_laser-driven_2002}.  We will show that the corresponding value of the QFI strongly depends upon the reflecting phase shift, and that for a coherent state the optimal phase shift corresponds to the presence of an anti-node at the location where the cavity couples to the fiber, see Figure~\ref{fig:schematic}.

For input NOON states, we find that the QFI is maximized not at an anti-node, but at a point of maximal slope in the interference pattern. Specifically, we will show below that for the NOON state with N photons, the optimized QFI at $\epsilon = 0$ obeys the bounds $64 N^2/\gamma^2\le \mathcal{I_Q}(0) \le 108 N^2/\gamma^2$, where the upper (lower) bound coincides with $\rho = 1$ ($\rho = 0$) respectively.  While the relative enhancement of $\mathcal{I_Q}(0)$ between the DP and EP ($108/64 \approx 1.69$ ) is smaller than that for a shot-noise limited coherent state, the resulting minimum MSE for an input NOON state scales with the number of photons as $N^{-2}$, corresponding to the $1/N$ Heisenberg scaling of the standard error $\sqrt{MSE_{\epsilon}}$.

The remainder of the paper is structured as follows. Section~\ref{sec:model} describes the quantitative system model and derives the linear response transfer function for the system.  Section~\ref{sec:fisher} calculates the QFI for a coherent state input and for a nonclassical NOON state, while leaving some details the appendices.  We also identify under what parameters the QFI is optimized in both cases.  From these optimal parameters, section~\ref{sec:detection} proposes two simple detection schemes, homodyne detection and a frequency-shifted photon counting interferometer, are able to achieve the QCRB for a coherent state and for a NOON state, respectively.  Finally, in section~\ref{sec:conclusion} we provide a discussion of the connection between an ES and the optimal QFI, indicating which non-Hermitian systems can be expected to yield divergent QFI, and also highlighting the role of coherent quantum feedback in this asymmetric system when driving with input quantum fields.  

\section{\label{sec:model}The model}
Our analysis is based on the formalism of input/output (IO) theory commonly used in quantum optics~\cite{loudon_quantum_2000}.  We assume for simplicity that the CW and CCW modes in the MRR have a single resonance frequency $\omega_{cav}$.  Furthermore, we work in a high-Q limit where each mode is modeled as a simple harmonic oscillator, with bosonic field operators $\hat{a}_{cw}$ and $\hat{a}_{ccw}$ ($[\hat{a}_{i}, \hat{a}_{j}^\dag ] = \delta_{ij}$ for $i,j \in \{ cw, ccw\}$).  The rate $\gamma$ sets the characteristic time scale for light to decay from the cavity and thus also characterizes the coupling strength between the inputs and the MRR.  The high-Q limit then specifies that $\gamma \ll \omega_{cav}$.  In a frame rotating at frequency $\omega_{cav}$, the Heisenberg equations of motion for the MRR satisfy: 
\begin{equation}\label{eq:cav_sources}
    \frac{d}{dt} \left[\begin{array}{c}
    \hat{a}_{cw}\\
    \hat{a}_{ccw}
    \end{array}\right] = - i\,\left[ \begin{matrix} -i \tfrac{\gamma}{2} & \epsilon \\ \epsilon  & -i \tfrac{\gamma}{2} \end{matrix}\right]\,\left[\begin{array}{c}
    \hat{a}_{cw}\\
    \hat{a}_{ccw}
    \end{array}\right] + \text{``sources''}. 
\end{equation}
The diagonal terms, $-i \gamma/2$, represent the decay of the cavity field into the coupling waveguide. The perturbing impurity induces the cross-coupling between the cavity field modes, which is parameterized by $\epsilon$. 

A detailed derivation of this equation of motion can be found in Appendix~\ref{app:IO_HLE} using the formalism of input-output theory.  Here we summarize the analysis with a physically motivated description.

The coupling waveguide is assumed to support a single spatial mode with both left and right propagating one-dimensional (1D) optical fields, see Figure~\ref{fig:schematic}.  We are interested in input states of light that are near resonance with the cavity and so we assume that the field inputs have a center carrier frequency equal to $\omega_{cav}$.  When the input states of light are restricted to a bandwidth $B \sim \gamma \ll \omega_{cav}$, we can model the 1D input fields as quantum white noise~\cite{gardiner_quantum_2004}. In a quantum white noise model,  the left (right) input field is modeled by delta-commuting bosonic operators $\hat{a}^{in}_l(t)$ ($\hat{a}^{in}_r(t)$), where $[\hat{a}^{in}_{i}(t), \hat{a}^{in\dag}_{j}(t') ] = \delta_{ij} \delta(t - t')$  for $i,j \in \{ l,r\}$.  We note that, in order for these input fields to have a commutation relation proportional to a delta function in time (which has units of sec$^{-1}$), then they must have units of sec$^{-1/2}$. 

The source terms in Eq.~(\ref{eq:cav_sources}) originate from a linear coupling between the cavity modes and the left/right input fields.  The source for the CW mode is simply proportional to the left-side input-field, $\hat{a}^{in}_l(t)$, with a proportionality constant of $-\sqrt{\gamma}$.  The reason for this particular proportionality constant is that 
the linear Heisenberg-Langevin equation (HLE),
\begin{equation}
    \tfrac{d}{dt} \hat{a}_{cw} = -\tfrac{\gamma}{2} \hat{a}_{cw} + g\, \hat{a}^{in}_l(t),
\end{equation}
with a complex coupling coefficient $g$,
has the solution
\begin{equation}\label{eq:CW_mode_g_input}
    \hat{a}_{cw}(t) = \hat{a}_{cw}(0)\, e^{-\gamma t/2} + g\, \int_0^t ds\, e^{-\gamma (t -s)/2}\, \hat{a}^{in}_l(s).
\end{equation}
The initial value of the cavity field, $\hat{a}_{cw}(0)$, is assumed to be statistically independent from the input fields and therefore commutes with $\hat{a}_{l}^{in}(t)$ and $\hat{a}_{l}^{in\,\dag}(t)$.
However, in order for the equal-time canonical commutation relations to be stationary, i.e., for
\begin{equation}
    [\hat{a}_{cw}(t), \hat{a}^\dag_{cw}(t)] = [\hat{a}_{cw}(0), \hat{a}_{cw}^\dag(0)] = 1,
\end{equation}
it must be true that $|g|^2 = \gamma$.  This can be seen by explicitly computing $[\hat{a}_{cw}(t), \hat{a}^\dag_{cw}(t)]$ from Eq.~(\ref{eq:CW_mode_g_input}) and its adjoint.  Our choice of $g = -\sqrt{\gamma}$, rather than some other complex phase value, is for convenience, since this particular choice simplifies the phase relationship between the input field and the field exiting the cavity.  This is further discussed in Appendix~\ref{app:IO_HLE}.

In light of this, the source term for the CCW mode will have two contributions.  The first is proportional to the transmitted part of $\hat{a}^{in}_r(t)$.  The second part is proportional to the field reflected from the left side of the beam splitter.  However, the key point is that this reflection has itself two parts.  The first is the reflection from $\hat{a}^{in}_l(t)$ and the second is the decaying output from the CW cavity mode.  Putting all of this together results in 
\begin{equation}\label{eq:sources}
    \text{``sources''} = -\sqrt{\gamma} \left[\begin{matrix}
        \hat{a}^{in}_l(t) \\
        \tau e^{i\phi}\, \hat{a}^{in}_r(t) + \rho e^{i 2 \phi}\big( \hat{a}^{in}_l(t) + \sqrt{\gamma}\, \hat{a}_{cw} \big)
    \end{matrix}
    \right],
\end{equation}
where $\phi$ ($2\phi$) is the phase change when transmitting (reflecting) across the system, respectively.  Combined into a matrix equation the vector of cavity field operators $\hat{\mathbf{a}}^{cav}(t) = [\hat{a}_{cw}(t), \hat{a}_{ccw}(t)]^T$ obeys the HLE:
\begin{equation}\label{eq:cav_HOM}
    \frac{d}{dt} \hat{\mathbf{a}}^{cav}(t) = -i\, \mathbf{\tilde{H}}_\epsilon \cdot \hat{\mathbf{a}}^{cav}(t)  - \mathbf{B}\cdot \hat{\mathbf{a}}^{in}(t).
\end{equation}
where the Hamiltonian $\mathbf{\tilde{H}}_\epsilon$ is given by
\begin{equation}\label{eq:H_decay}
    \mathbf{\tilde{H}}_\epsilon = \left[\begin{matrix}
        -i \tfrac{\gamma}{2} & \epsilon\\
         \epsilon - i \gamma\, \rho\, e^{i 2 \phi} & - i \tfrac{\gamma}{2}
        \end{matrix}\right]
\end{equation}
and the transformation matrix $\mathbf{B}$ is given by
\begin{equation}\label{eq:B}
   \mathbf{B} =  \sqrt{\gamma}\left[\begin{matrix}
                    1 & 0\\
                    \rho\, e^{i 2 \phi} & \tau\, e^{i \phi}
                \end{matrix}\right].  
\end{equation}

The eigenvalues of $\mathbf{\tilde{H}}_\epsilon$ are
\begin{equation}\label{eq:Htilde_eigvals}
    \Omega_{\pm} = -i \tfrac{\gamma}{2} \pm \sqrt{\epsilon^2 - i \epsilon \gamma \rho e^{i 2 \phi} },
\end{equation}
which become degenerate when $\epsilon = 0$.  In addition to being degenerate, for $\rho > 0$ the two eigenvectors of $\mathbf{\tilde{H}}_{0} \equiv \mathbf{\tilde{H}}_{\epsilon = 0}$ coalesce and is an EP.  When viewed as a function of the remaining parameters $(\gamma, \rho, \phi)$, $\mathbf{\tilde{H}}_0$ will remain singular, so that the continuous set of all parameterizations with $\epsilon = 0$ comprises an exceptional surface.  

The general solution to Eq.~(\ref{eq:cav_HOM}) is
\begin{equation}\label{eq:cav_sol_t}
\begin{split}    
    \hat{\mathbf{a}}^{cav}(t) &= \exp\left[-i \mathbf{\tilde{H}}_\epsilon (t-t_0)\right] \cdot \hat{\mathbf{a}}^{cav}(t_0) \\
    &\quad - \int_{t_0}^t ds\, \exp\left[-i \mathbf{\tilde{H}}_\epsilon (t-s)\, \right] \cdot \mathbf{B}\cdot \hat{\mathbf{a}}^{in}(s) .
\end{split}
\end{equation}
We assume that the initial time $t_0$ is in the distance past and take the limit $t_0 \rightarrow -\infty$ so that  any effect of the initial condition will have decayed away.  From here on, we ignore the initial value term and focus solely on the second source term of Eq.~(\ref{eq:cav_sol_t}).

In IO theory, both the system (cavity) and the input fields $\hat{a}^{in}_i(t)$ evolve under a joint unitary $U(t)$.  To ensure a unitary description, we assume that all losses can be modeled by additional IO modes.  However, any light lost to an unmeasured environment will reduce the amount of information gained.  Therefore, in order to find an optimum bound, we focus on the ideal lossless system.  For the vector of inputs $\hat{\mathbf{a}}^{in}(t) = \left[ \hat{a}^{in}_l(t), \hat{a}^{in}_r(t)\right]^T$, the Heisenberg evolution under $U(t)$ results in a unitary transformation between the input and output modes.  In general, we have $\hat{\mathbf{a}}^{out}(t) = U^{\dag}(t)\, \hat{\mathbf{a}}^{in}(t)\, U(t)$.  All estimates of $\epsilon$ are obtained from measurements of the output light, i.e., from the statistics of observables constructed from $\hat{\mathbf{a}}^{out}(t)$ and $\hat{\mathbf{a}}^{out\dag}(t)$. 

In the absence of the cavity (e.g. $\gamma = 0$), the IO relation is given by a simple beam splitter transformation
\begin{equation}
        \hat{\mathbf{a}}^{out}(t) = \mathbf{S} \cdot \hat{\mathbf{a}}^{in}(t)
\end{equation}
where
\begin{equation}\label{eq:S_matrix}
    \mathbf{S} = \left[ \begin{array}{cc}
            \rho\, e^{i 2\phi} & \tau\, e^{i \phi}\\
            \tau\, e^{i \phi} & -\rho
        \end{array}\right]
\end{equation}
and is unitary.  The left-side scattering phase shift $\phi$ is included in $\mathbf{S}$ to explicitly account for the optical path length between the resonator and the reflecting beam splitter.  We assume that the light travel time across this length is negligible when compared to the cavity decay time $1/\gamma$.  

For a non-zero cavity coupling $\gamma > 0$, the output fields are given by the superposed sum of the beam splitter transformed input fields ($\mathbf{S} \cdot \hat{\mathbf{a}}^{in}(t)$ ) with the sources from the cavity.  In particular, the right-side output has a contribution from the CW mode, after transmitting through the BS.  However, the left-side output contains all of the light from the CCW mode, as well as the CW reflection from the BS.  Combining this results in the IO relation, 
\begin{equation}\label{eq:io_t}
\begin{split}
    \hat{\mathbf{a}}^{out}(t) &= \mathbf{S} \cdot \hat{\mathbf{a}}^{in}(t) + \sqrt{\gamma} \left[\begin{matrix}
        \rho\, e^{i2\phi} & 1 \\
        \tau\, e^{i\phi} & 0 
    \end{matrix} \right] \cdot \hat{\mathbf{a}}^{cav}(t)\\
    &= \mathbf{S} \cdot \hat{\mathbf{a}}^{in}(t) + \mathbf{S}\cdot \mathbf{B}^\dag \cdot \hat{\mathbf{a}}^{cav}(t)
\end{split}
\end{equation}
While it is not immediately obvious that the second line follows from the first, it is easily verified by computing the product $\mathbf{S\cdot B}^\dag$ and comparing terms.

Since the cavity fields $\hat{\mathbf{a}}^{cav}(t)$ from Eq.~(\ref{eq:cav_sol_t}) are given by a linear response integral over the past input fields, this solution is best expressed in the frequency domain where the transformation is given by a local point-wise transfer function of frequency.  Upon taking the Fourier transform of Eq.~(\ref{eq:io_t}), (and ignoring the initial value term) the IO relation is
\begin{equation}\label{eq:io_w}
\begin{split}
    \mathbf{\hat{a}}^{out}(\omega) &= \left(\mathbf{S} -\mathbf{S}\mathbf{B}^\dag \cdot \frac{1}{i \mathbf{\tilde{H}}_\epsilon - i \omega} \cdot \mathbf{B} \right )\cdot \mathbf{\hat{a}}^{in}(\omega)\\
    \mathbf{\hat{a}}^{out}(\omega) &\equiv \mathbf{K}_\epsilon(\omega)\cdot \mathbf{\hat{a}}^{in}(\omega),
\end{split}    
\end{equation}
with the transfer function $\mathbf{K}_\epsilon(\omega)$ defined as
\begin{equation}\label{eq:K(w)}
\begin{split}
    \mathbf{K}_\epsilon(\omega) &= \mathbf{S}\cdot\left(\mathbf{I} + i \mathbf{B}^\dag\cdot \mathbf{R}_{\epsilon}(\omega )\cdot\mathbf{B}\, \right).
\end{split}
\end{equation}
Here $\mathbf{I}$ is the $2\times 2$ identity matrix and $\mathbf{R}_\epsilon(\omega)$ is the resolvent of $\mathbf{H}_\epsilon$, defined as $\mathbf{R}_\epsilon(\omega) \equiv (\tilde{\mathbf{H}}_\epsilon - \omega \mathbf{I} )^{-1}$.
In our system, $\mathbf{K}_\epsilon$ has the matrix elements,
\begin{subequations}
\begin{align}
 K_{ll} &= -i\gamma\frac{\epsilon(1 + \rho^2 e^{i 4 \phi} ) + 2 \rho e^{i 2\phi} \omega }{(\omega - \Omega_+)(\omega - \Omega_-)} + \rho e^{i 2\phi}\label{eq:Kll} \\
 K_{lr} &= -i\gamma \tau e^{i \phi} \frac{\omega + i \tfrac{\gamma}{2} + \epsilon\rho e^{i 2 \phi} }{(\omega - \Omega_+)(\omega - \Omega_-)}  + \tau e^{i\phi}\\
 K_{rr} &= -i\gamma\frac{\epsilon\, \tau^2 e^{i 2 \phi}}{(\omega - \Omega_+)(\omega - \Omega_-)} - \rho\\ 
 K_{rl} &= K_{lr}, 
\end{align}
\end{subequations}
where $\Omega_{\pm}$ are the complex eigenvalues of $\tilde{H}_{\epsilon}$ given in Eq.~(\ref{eq:Htilde_eigvals}). These matrix elements were computed using the symbolic mathematics library SymPy~\cite{meurer_sympy_symbolic_computing_2017}.

While it is not apparent from the above matrix elements, it can be confirmed that $\mathbf{K}(\omega)$ is a unitary matrix.  A detailed discussion is given in App.~\ref{app:A}, where we show that a necessary and sufficient condition for unitarity is that $\operatorname{Im} \tilde{\mathbf{H}_\epsilon} = -\frac{1}{2} \mathbf{BB}^\dag$, which is the case for this system.  Then using Eq.~(\ref{eq:io_w}) we will find that the commutation relations for the input fields are correctly preserved when transforming to the outputs, e.g., $[\hat{a}_l^{out}(\omega_1), \hat{a}_r^{out\dag}(\omega_2)] = [\hat{a}_l^{in}(\omega_1), \hat{a}_r^{in\dag}(\omega_2)] = 0$.

\section{Fisher information \label{sec:fisher} }
When discussing the precision of sensing a particular parameter $\epsilon$, a useful quantifying metric is the mean squared error (MSE) of a specific estimator for that parameter. We denote the estimator for $\epsilon$ as $\hat\epsilon$.  The utility of this particular metric is that the (classical) Cram\'er-Rao bound sets a lower bound on the MSE and depends only on the underlying probability distribution $p(x|\epsilon)$, given no statistical biases in $\hat{\epsilon}$.  

In the quantum setting, the model is traditionally formulated as a set of density matrices $\{\rho_{\epsilon} : \epsilon \in \mathcal{E}\}$ for an index set of valid parameter values $\mathcal{E} \subseteq \mathbb{R}$.  
The general framework for modeling a quantum measurement whose outcomes $x$ are drawn from the set $\mathcal{X}$, is to specify the positive operator-valued measure (POVM) with elements $\{E_x : x\in \mathcal{X}\}$. When given the state $\rho_\epsilon$ the probability (density) for outcome $x$ is $p(x|\epsilon) = \operatorname{tr}[E_x \rho_\epsilon ]$.  
In order for the POVM to define a valid probability measure for any state $\rho$, the elements $E_x$ must be positive operators (i.e. $E_x \ge 0$) and satisfy the constraint $\sum_{x\in \mathcal{X}} E_x = \mathbf{I}$.
Our estimator is then a (classical) function $\hat{\epsilon} : \mathcal{X} \rightarrow \mathcal{E}$, which returns an estimate $\epsilon_{est}$ given a measurement outcome $x$.  

Given $\{E_x\}$ and an estimator $\hat{\epsilon}$, this estimator is called unbiased if, for all parameter values $\epsilon \in \mathcal{E}$,
\begin{equation}
 \mathbf{E}_{\epsilon}(\hat{\epsilon}) \equiv \sum_{x \in \mathcal{X}} \hat{\epsilon}(x)\, \operatorname{tr}[E_x \rho_\epsilon ] = \epsilon,
\end{equation} 
i.e., the estimator performs correctly on average.
When $\hat\epsilon$ is unbiased, the single trial MSE is equal its variance,
\begin{equation}
    \text{MSE}_\epsilon \equiv \sum_{x \in \mathcal{X}} \left(\hat{\epsilon}(x)^2 - \epsilon^2\right) \operatorname{tr}(E_x\, \rho_\epsilon) = \mathbb{V}_{\epsilon}(\hat{\epsilon}).
\end{equation}
The QCRB states that after $m$ independent trials, the MSE of an unbiased estimator satisfies the inequality\cite{braunstein_statistical_1994,ye_quantum_2022}  
\begin{equation}
    \mathbb{V}_{\epsilon}(\hat{\epsilon}) \ge\frac{1}{m\, \mathcal{I_{Q}}(\epsilon)},
\end{equation}
where the QFI, $\mathcal{I_Q}$, is independent of $\hat{\epsilon}$ and depends only on $\rho_{\epsilon}$ and $d \rho_\epsilon/d\epsilon$.  Since the number of trials only enters the discussion as a multiplicative factor, we will now set $m = 1$ and focus solely on the QFI.

Formally, the QFI is given by $\mathcal{I_Q}(\epsilon) \equiv \operatorname{tr}[\rho_{\epsilon} L_\epsilon^2]$ where $L_\epsilon$ is the symmetric logarithmic derivative of $\rho_\epsilon$.  Subtleties arise when $L_\epsilon$ becomes unbounded and result in discontinuities in $\mathcal{I_Q}(\epsilon)$~\cite{safranek_discontinuities_2017}, which generally occur when the rank of $\rho_{\epsilon}$ changes.  

When $L_\epsilon$ is bounded, the Fisher information $\mathcal{I_Q}(\epsilon)$ is continuous and can be expressed as the limit of the Bures distance between the state $\rho_\epsilon$ and $\rho_{\epsilon + \delta\epsilon}$, 
\begin{equation}
    d_{B}(\rho_{\epsilon}, \rho_{\epsilon + \delta\epsilon})^2 = 2 \left( 1- \sqrt{\mathcal{F}(\rho_{\epsilon}, \rho_{\epsilon + \delta\epsilon}) } \right)
\end{equation}
where the fidelity is 
\begin{equation}
    \mathcal{F}(\rho, \sigma) = \operatorname{tr}\left[ \sqrt{\sqrt{\sigma}\rho \sqrt{\sigma }}\right]^2. 
\end{equation}
Given this, the QFI is then~\cite{ye_quantum_2022}
\begin{equation}\label{eq:QFI_bures}
    \mathcal{I}_{Q}(\epsilon) = 4 \lim_{\delta\epsilon \rightarrow 0} \frac{ d_B(\rho_{\epsilon}, \rho_{\epsilon + \delta\epsilon})^2}{\delta\epsilon^2}.
\end{equation}
This expression will ultimately be more convenient for computing the QFI directly than evaluating the expectation value of the square of the symmetric logarithmic derivative, $\operatorname{tr}[\rho_{\epsilon} L_\epsilon^2]$.  We show below that it is relatively straightforward to calculate the fidelity between two possible output fields for our system and hence to find the limit defined in Eq.~(\ref{eq:QFI_bures}).  This is done below for two possible input states: a coherent state of amplitude $\boldsymbol{\beta}$ and a nonclassical NOON state.  In order to lighten the notation and easily compare the QFI given these two states we will denote $\mathcal{I}_{Q}$ for a coherent state as $\mathcal{I}_{\beta}$ and for a NOON state QFI will be denoted as $\mathcal{I_N}$.

\subsection{Coherent state inputs}

If we have a vector of coherent input amplitudes, $\boldsymbol{\beta}^{in}(\omega) = [\beta_l(\omega), \beta_r(\omega)]^T$, and the fact that a coherent state $|\psi[\boldsymbol{\beta}^{in}]\rangle$ is an eigenstate of $\hat{a}_i(\omega)$ with eigenvalue $\beta_i(\omega)$ we then have the state transformation
\begin{equation}
    |\psi[\boldsymbol{\beta}^{out}] \rangle = |\psi[\mathbf{K}\cdot\boldsymbol{\beta}^{in}] \rangle.
\end{equation}

The fidelity between two pure coherent states with amplitudes $\boldsymbol{\beta}$ and $\boldsymbol{\beta}'$  
is
\begin{equation}
    |\langle \psi[\boldsymbol{\beta}']|\, \psi[\boldsymbol{\beta}]\rangle|^2 = \exp\left( -\|\boldsymbol{\beta}'\|^2 - \|\boldsymbol{\beta}\|^2 + 2\operatorname{Re} \langle \boldsymbol{\beta}', \boldsymbol{\beta}\rangle \right)
\end{equation}
where we have introduced the inner product in the space of coherent state wavepackets~\cite{garrison_quantum_optics_2008}
\begin{equation}
    \langle\boldsymbol{\beta}', \boldsymbol{\beta} \rangle = \sum_i \int d\omega \  \beta^{'*}_i(\omega) \beta_i(\omega).
\end{equation}

Defining
\begin{equation}\label{eq:beta_eps}
    \boldsymbol{\beta}_{\epsilon}(\omega) \equiv \mathbf{K}_{\epsilon}(\omega) \cdot \boldsymbol{\beta}(\omega)
\end{equation}
the fidelity becomes
\begin{equation}
\begin{split}
    |\langle \psi[\boldsymbol{\beta}_{\epsilon}]| \psi[\boldsymbol{\beta}_{\epsilon +\delta\epsilon}]\rangle|^2 &= \exp\left( -2 \|\boldsymbol{\beta}\|^2 + 2\operatorname{Re}\, \langle \boldsymbol{\beta}_{\epsilon}, \boldsymbol{\beta}_{\epsilon +\delta\epsilon}\rangle \right)\\
    =& \exp\left( -2 \|\boldsymbol{\beta}\|^2 + 2\operatorname{Re}\, \langle \mathbf{K}_\epsilon \boldsymbol{\beta},  \mathbf{K}_{\epsilon + \delta\epsilon} \boldsymbol{\beta}\rangle \right)\\
    =& \exp\left( -2 \|\boldsymbol{\beta}\|^2 + 2\operatorname{Re}\, \langle \boldsymbol{\beta}, \mathbf{K}_\epsilon^\dag \mathbf{K}_{\epsilon + \delta\epsilon} \boldsymbol{\beta}\rangle \right).
\end{split}
\end{equation}
Expanding this equation to second order in $\delta\epsilon$ results in a straightforward calculation of the QFI in Eq.~(\ref{eq:QFI_bures}).

In Appendix~\ref{app:A}, we show that when the Hamiltonian $\mathbf{\tilde H}_\epsilon = \mathbf{\tilde{H}_{0}} + \epsilon \mathbf{V}$ is subject to the perturbation $\epsilon \mapsto \epsilon + \delta\epsilon$ we have the relation, 
\begin{multline} \label{eq:KdK_in_A}
    \mathbf{K}_{\epsilon}^\dag \mathbf{K}_{\epsilon + \delta\epsilon} = \mathbf{I} + i\, \mathbf{A}_{\epsilon}\,\delta\epsilon\\ + \left(- \mathbf{A}_{\epsilon}^2 +i \frac{\partial \mathbf{A}_{\epsilon}}{\partial \epsilon} \right) \frac{\delta\epsilon^2}{2} + O(\delta\epsilon^3),
\end{multline}
where $\mathbf{A}_{\epsilon}$ is the Hermitian generator for local changes in $\epsilon$, which is defined as
\begin{equation}\label{eq:A_as_GWSM}
\begin{split}
    \mathbf{A}_{\epsilon} &= -\mathbf{B}^\dag\mathbf{R}_{\epsilon}^\dag\, \mathbf{V}\,\mathbf{R}_{\epsilon}\, \mathbf{B}\\
    &= -i \mathbf{K}^\dag_\epsilon \frac{\partial\mathbf{K}_{\epsilon}}{\partial \epsilon}.
\end{split}
\end{equation}
Here the second line follows from Eq.~(\ref{eq:KdK_in_A}).  This matrix plays a special role in optical scattering theory, where it is called a generalized Wigner-Smith matrix (GWSM)~\cite{rachbauer_how_2024}.  Were this to contain a derivative with respect to $\omega$, rather than $\epsilon$, then $\mathbf{A}$ would be the Wigner-Smith matrix whose eigenvalues are the ``proper time delays'' in the scattering process. Rather than determining how the scattering phase shift changes with frequency, our GWSM, $\mathbf{A}_\epsilon$, is the matrix describing the response of $\mathbf{K}$ to an infinitesimal change in $\epsilon$.  However as momentum generates translations in its conjugate variable position, $\mathbf{A}_\epsilon$ is a generalized momentum conjugate to $\epsilon$.

Returning to the fidelity, we note that this only depends upon the real part of the inner product $\left\langle \boldsymbol{\beta},\, \mathbf{K}^\dag_{\epsilon} \mathbf{K}_{\epsilon + \delta\epsilon}\, \boldsymbol{\beta} \right\rangle$, so that the output coherent state fidelity is equal to
\begin{equation}
    \mathcal{F}(\boldsymbol{\beta}_{\epsilon}, \boldsymbol{\beta}_{\epsilon + \delta\epsilon} ) = 1 - \langle \boldsymbol{\beta}, \mathbf{A}_\epsilon^2 \boldsymbol{\beta}\rangle\, \delta\epsilon^2 + O(\delta\epsilon^3).
\end{equation}
This results in a coherent state QFI of
\begin{equation}
    \mathcal{I}_{\beta}(\epsilon) = 4\, \langle \boldsymbol{\beta}, \mathbf{A}_\epsilon^2 \boldsymbol{\beta}\rangle,
\end{equation}
which is maximized when $\boldsymbol{\beta}$ is proportional to the eigenstate of $\mathbf{A}_{\epsilon}^2$ with a maximal eigenvalue $\lambda_{abs}^2$.  Thus we have the bound
\begin{equation}\label{eq:QFI_cs}
    \mathcal{I}_{\beta}(\epsilon) \le 4 \lambda_{abs}^2\, \|\boldsymbol{\beta}\|^2.
\end{equation}

Note that because $\mathbf{A}_\epsilon$ is a matrix whose entries are functions of frequency, it can be viewed as an operator that is diagonal in the frequency domain.  Thus its eigenstates will be single mode plane waves and are subject to a delta-function normalization condition.  We note that, in practice, the actual input state will be some smooth normalizable wave packet, e.g. a Gaussian, so that any frequency optimized quantities we calculate are technically upper bounds on the QFI.

\subsection{NOON state inputs}
It is known that in general, for bounded resources certain entangled states are able to achieve a larger Fisher information than non-entangled classical states~\cite{giovannetti_quantum_2006}.  Furthermore, when changes in the parameter $\epsilon$ are given by a unitary transformation $U(\epsilon) = \exp(i \epsilon G)$, then the optimal state a superposition where all $N$ are either in the maximal or in the minimal eigenstate of the generator $G$ - a so-called NOON state when the given eigenstates are for two distinct optical modes

In our continuous mode context we will construct our NOON state out of orthogonal temporal and/or spatial modes propagating in the optical network.  Mathematically, we will consider two mode functions $\boldsymbol{\psi}_1(\omega)$ and $\boldsymbol{\psi}_2(\omega)$ such that $\langle\boldsymbol{\psi}_n, \boldsymbol{\psi}_m \rangle = \delta_{nm}$.  We then have the frequency-smeared field operators $\hat{a}^\dag[\boldsymbol{\psi}_n] \equiv \sum_n \int d\omega\ \mathbf{\hat{a}}^{in\, \dag}(\omega) \cdot \boldsymbol{\psi}_{n}(\omega)$.  (Note that $\left[\hat{a}[\boldsymbol{\psi}_n], \hat{a}^\dag[\boldsymbol{\psi}_m]\right] = \delta_{nm}$.)
Finally our NOON input state is  
\begin{equation}
    |\Psi \rangle  = \frac{1}{\sqrt{2 N!}} \sum_{n = 1}^2 \left.\hat{a}^\dag[\boldsymbol{\psi}_n] \right.^N |vac\rangle,
\end{equation}
where $|vac \rangle$ is the zero photon vacuum state.)

The statistics of the outgoing light is calculated via the input-output transformation of Eq.~(\ref{eq:io_w}).  For example, the $N$-photon Fock state $| \boldsymbol{\psi}_N \rangle$ has the left-going output photon-number spectral density
\begin{equation}
\begin{split}
    \langle  \boldsymbol{\psi}_N| \hat{a}^{out\,\dag}_l(\omega) \hat{a}^{out}_l(\omega) | \boldsymbol{\psi}_N\rangle = \left\|\hat{a}^{out}_l(\omega)\, |\boldsymbol{\psi}_N \rangle \right\|^2\\ 
    = \left\|\sqrt{N} \, \mathbf{e}_l\cdot \mathbf{K}(\omega)\cdot \boldsymbol{\psi}(\omega) | \boldsymbol{\psi}_{N-1}\rangle\right\|^2\\ 
    \equiv N \left|\psi^{out}_l(\omega)\right|^2,
\end{split}
\end{equation}    
where in the output Schrodinger picture, we have defined the single photon amplitude $\boldsymbol{\psi}^{out}(\omega)$.
\begin{equation}
    \boldsymbol{\psi}^{out}(\omega) = \mathbf{K}_\epsilon(\omega)\cdot\boldsymbol{\psi}^{in}(\omega).
\end{equation}
Thus at the value of $\epsilon$, the output NOON state is 
\begin{multline}
    |\Psi_\epsilon \rangle = \\
    \frac{1}{\sqrt{2 N!}} \sum_{n = 1}^2 \left(\int d\omega\ \mathbf{\hat{a}}^{\dag}(\omega) \cdot \boldsymbol{\psi_n}^{out}(\omega) \right)^N |vac \rangle. 
\end{multline}

Calculating the NOON state fidelity, $|\langle \Psi_{\epsilon}|\Psi_{\epsilon + \delta \epsilon} \rangle|^2$, begins with implementing Wick's theorem to calculate 
\begin{equation}
\begin{split} \label{eq:NOON_wick}
    \langle \Psi_{\epsilon} | \Psi_{\epsilon + \delta\epsilon} \rangle &= \frac{N!}{2 N!} \sum_{n,m = 1}^2 \left\langle \mathbf{K}_{\epsilon}\cdot \boldsymbol{\psi_n}, \mathbf{K}_{\epsilon + \delta\epsilon}\cdot \boldsymbol{\psi_m} \right\rangle^N \\
    &= \frac{1}{2} \sum_{n,m = 1}^2 \left\langle \boldsymbol{\psi_n}, \mathbf{K}_{\epsilon}^\dag\mathbf{K}_{\epsilon + \delta\epsilon} \boldsymbol{\psi_m} \right\rangle^N
\end{split}
\end{equation}

In Appendix~\ref{app:NOON} we expand $|\langle \Psi_{\epsilon} | \Psi_{\epsilon + \delta\epsilon} \rangle|^2$ out to second order in $\delta \epsilon$.  We then write this fidelity in terms of an N-fold tensor product of ``single particle wave functions'' $\boldsymbol{\psi}_n$; $|NOON \rangle = \left(|\boldsymbol{\psi}^{\otimes N}_1\rangle + |\boldsymbol{\psi}^{\otimes N}_2 \rangle \right)/\sqrt{2} $ and the expectation value of a collective operator $\mathbf{A}^{tot}_\epsilon = \sum_{i=1}^{N} \mathbf{A}_\epsilon^{(n)}$.  The resulting form of the fidelity is ultimately given by
\begin{equation}
    \mathcal{F}(\epsilon, \epsilon + \delta\epsilon) = 1 - \delta\epsilon^2 \left( \Delta \mathbf{A}^{tot}_\epsilon \right)^2  + O(\delta\epsilon^3),
\end{equation}
where $\Delta \mathbf{A}^{tot}_\epsilon$ is the variance of $\mathbf{A}^{tot}_\epsilon$ under the input NOON state.
The QFI given a NOON state is then
\begin{equation}
    \mathcal{I_N}(\epsilon) = 4 \left( \Delta \mathbf{A}^{tot}_\epsilon \right)^2. 
\end{equation}
If $\mathbf{A}_\epsilon$ is bounded such that $\lambda_{\min} \le \mathbf{A}_{\epsilon} \le \lambda_{\max}$, we then have the following upper bound on the variance \cite{boixo_generalized_2007}
\begin{equation}
    \left( \Delta \mathbf{A}^{tot}_\epsilon \right)^2 \le \frac{N^2}{4}(\lambda_{\max} - \lambda_{\min} )^2, 
\end{equation}
which leads to the upper bound 
\begin{equation}\label{eq:QFI_noon}
    \mathcal{I_N}(\epsilon) \le N^2\, (\lambda_{\max} - \lambda_{\min})^2. 
\end{equation}

This bound is saturated by a NOON state composed of the equivalent eigenstates $\boldsymbol{\psi}_1 = |\lambda_{\min} \rangle $ and $\boldsymbol{\psi}_2 = |\lambda_{\max} \rangle$.

\subsection{Optimization of the Fisher information}
For both of the states we have considered in this work, the QFI is ultimately constrained by the eigenvalues of $\mathbf{A}_\epsilon$, namely $\lambda_{\pm}(\omega)$, which depend on the system parameters $\epsilon$, $\rho$, $\phi$, and $\gamma$.  We will ignore the $\gamma$ dependence, as its only role is to set the characteristic interaction time scale, which will effectively be set to 1.  For a given parameter set, we will then optimize the input state by maximizing the QFI for a known nominal value of $\epsilon$, e.g. $\epsilon = 0$.  This point-wise optimization results in input states being either a single monochromatic plane wave or a superposition of two plane waves, depending upon the choice of coherent or NOON input state.  In principle, one could construct a polychromatic input state, one that is optimized for a particular prior distribution over $\epsilon$.  We will leave this form of Bayesian region estimators for future work.  For brevity, we will call this frequency-optimized QFI the o-QFI.

In an abundance of clarity we define the frequency optimized values:
\begin{subequations}
\begin{align}
    \lambda_{\max} =& \max_{\omega \in \mathbb{R} } \lambda_{+}(\omega)\\
    \lambda_{\min} =& \min_{\omega \in \mathbb{R} } \lambda_{-}(\omega)\\
    \lambda_{\text{abs}} =& \max\, \{ |\lambda_{\min}|,\, |\lambda_{\max}| \},
\end{align}
\end{subequations}
These extrema generally occur at different values of $\omega$, which will also vary for different parameter values.  Therefore, $\lambda_{\max}$ and its maximizing frequency $\omega_{\max} \equiv \argmax_{\omega} \lambda(\omega)$ remain implicit functions of $\epsilon$, $\phi$, and $\rho$. The same is true for $\lambda_{\min}$ and the corresponding frequency $\omega_{\min}$.  While we will in principle have different values of the o-QFI for different system parameter values, we will also like to identify for what parameter values the o-QFI reaches a global maximum.  In particular, by comparing the o-QFI for different values of $\rho$, we are able to compare the optimal sensitivity of the $\epsilon = 0$ ES to the corresponding DP.  In order to perform this comparison, we now explicitly calculate the matrix elements of $\mathbf{A}_\epsilon$.  

Indexing the rows and columns by the input mode labels $l$ and $r$, symbolic evaluation of $\mathbf{A}_\epsilon$ results in
\begin{equation}
    \mathbf{A}_\epsilon = \left[\begin{matrix}
    A_{ll} & A_{lr} \\
    A_{rl} & A_{rr}
    \end{matrix} \right]
\end{equation}
with the matrix elements,
\begin{subequations}\label{eq:A_ij}
\begin{align}
    A_{ll} &= \frac{-\gamma\left( \epsilon(1 + \rho^2) \zeta + \rho e^{-i 2 \phi} (\zeta^2 + \epsilon^2 )\right) }{\left|\zeta^2 - \epsilon\left(\epsilon - i \gamma \rho e^{i2\phi}\right)\right|^2} + c.c.\\
    A_{rr} &= \frac{-\gamma\epsilon\tau^2( \zeta + \zeta^*)}{\left|\zeta^2 - \epsilon\left(\epsilon - i \gamma \rho e^{i2\phi}\right)\right|^2}\\
    A_{lr} &= A_{rl}^* = \frac{-\gamma\tau e^{i \phi} (\epsilon^2 + 2 \epsilon \rho e^{-i 2 \phi}\, \zeta + |\zeta|^2 )  }{\left|\zeta^2 - \epsilon\left(\epsilon - i \gamma \rho e^{i2\phi}\right)\right|^2},
\end{align}
\end{subequations}
where $\zeta = \omega + i \gamma/2$ and \emph{c.c.} stands for the complex conjugate of the preceding term.  We note that the denominator for each $A_{ij}$ is equal to $|\operatorname{det}(\tilde{\mathbf{H}}_\epsilon - \omega \mathbf{I} )|^2$.

From the above matrix elements $A_{ll}$ and $A_{rr}$ and noting that $\rho^2+\tau^2=1$, the trace of $\mathbf{A}_\epsilon$ simplifies to
\begin{equation}
    \operatorname{tr} ( \mathbf{A}_\epsilon ) =  \frac{-\gamma\left( 2 \epsilon \zeta + \rho e^{-i 2 \phi} (\zeta^2 + \epsilon^2 )\right) }{\left|\zeta^2 - \epsilon\left(\epsilon - i \rho e^{i2\phi}\right)\right|^2} + c.c.\,.
\end{equation}
A far less trivial calculation results in
\begin{equation}
        \operatorname{det} ( \mathbf{A}_\epsilon ) = \frac{ - \gamma^2\, \tau^2}{\left|\zeta^2 - \epsilon\left(\epsilon - i \rho e^{i2\phi}\right)\right|^2},
    \end{equation}
which we have again computed symbolically.  
In terms of these quantities, the eigenvalues are $\lambda^{(\epsilon)}_{\pm} = \operatorname{tr}(\mathbf{A}_\epsilon)/2 \pm \sqrt{\operatorname{tr}(\mathbf{A}_\epsilon)^2 - 4 \operatorname{det} ( \mathbf{A}_\epsilon ) }/2$.  For the special case of $\epsilon = 0$ this evaluates to
\begin{multline}\label{eq:la0_of_z}
    \lambda^{(0)}_{\mp}(\omega) =  - \frac{\gamma}{|\zeta|^2}\Big( \rho  \operatorname{Re} (  e^{i 2\phi} \zeta^*/\zeta ) \\
    \pm  \sqrt{ \tau^2 + \rho^2 \,\operatorname{Re}( \zeta^* e^{i 2 \phi} /\zeta )^{2} } \Big),
\end{multline}
which we denoted as a function of $\omega$, via the frequency dependence of $\zeta$.  Since $|\operatorname{Re} (  e^{i 2\phi} \zeta^*/\zeta )| \le 1$, we have the bound
\begin{equation}
    \lambda^{(0)}_{+} \le  \frac{\gamma}{|\zeta|^2} (1 + \rho) \le \frac{4}{\gamma} (1 + \rho),
\end{equation}
which saturates at $\omega = 0$ and $\phi = n\,\pi$ for integer $n$.  (Likewise, we have $ \lambda^{(0)}_{-} = - \frac{4}{\gamma}(1 + \rho)$ for $\omega = 0$ and $\phi = \pi\, ( n + 1/2) $.)  Thus regardless of the sign of  $ \lambda^{(0)}_{\pm}(\omega)$, we find that $ \lambda^{(0)}_{\pm}(\omega)^2 \le 16 (1 + \rho)^2/\gamma^2$.
so that at $\epsilon = 0$, the coherent state QFI has the phase-independent upper bound
\begin{equation} \label{eq:max_QFI_cs}
    \mathcal{I}_{\beta}(0) \le 64\, (1 + \rho)^2 |\boldsymbol{\beta}|^2/\gamma^2 \le 256\,|\boldsymbol{\beta}|^2/\gamma^2.
\end{equation}
This unambiguously shows that an input coherent state has a larger sensitivity on the $\epsilon = 0$ ES, when compared to the corresponding DP at $\rho=0$.
\begin{figure}[tbh]
    \includegraphics[width=1.0\columnwidth]{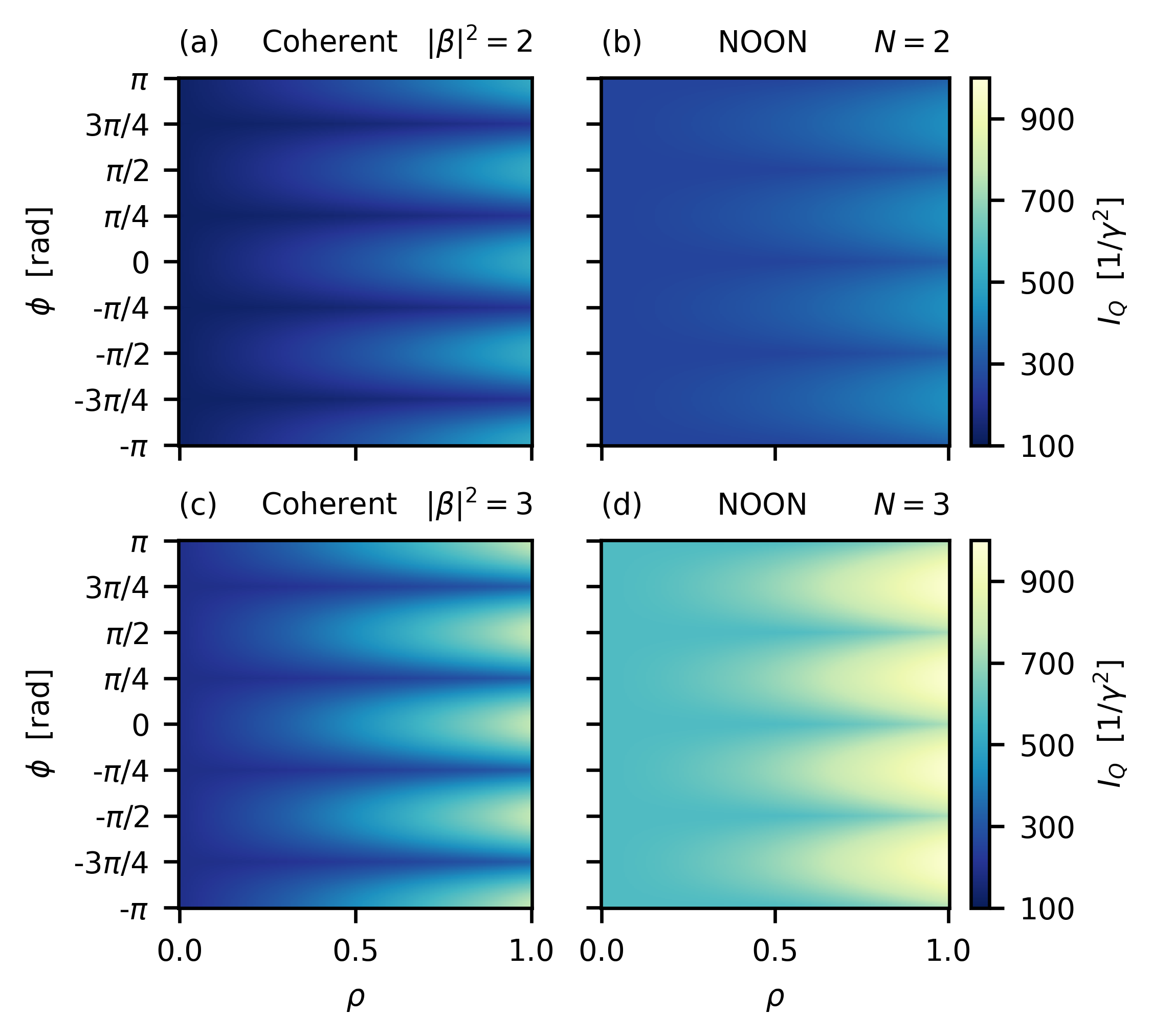}
    \caption{\label{fig:QFI} [color online] Frequency-optimized quantum Fisher information (o-QFI) on the ES $\epsilon = 0$.  This figure shows the o-QFI for a coherent state $\mathcal{I}_\beta$ (left column, panels (a) and (c)) and for a NOON state $\mathcal{I_N}$ (right column, panels (b) and (d)) as a function of the reflectivity $\rho$ and the reflection phase $\phi$.  The ES is defined for all values of $\phi$ and $\rho > 0$.  The first row, panels (a) and (b), shows results for $|\boldsymbol{\beta}|^2 = N = 2$, while the second row, panels (c) and (d), shows results for $|\boldsymbol{\beta}|^2 = N = 3$. For both input states, the peak QFI occurs at $\rho = 1$ and is periodic in $\phi$ where the coherent (NOON) state is maximal at even (odd) multiples of $\pi/4$, respectively. The maximal contrast is exactly a factor of 4 for a coherent state, a factor of $108/64 = 1.6875$ for a NOON state, and both are independent of the number of photons. The crossover between shot-noise limited and Heisenberg scaling occurs at $N = 3$, which shows a larger o-QFI for the NOON state.} 
\end{figure}
Fig.~\ref{fig:QFI} (a) and (c) plots the coherent state o-QFI for $\epsilon = 0$, as a function of $\rho$ and $\phi$ for an average photon numbers of $|\boldsymbol{\beta}|^2 = 2$ and $|\boldsymbol{\beta}|^2 = 3$, respectively.  The fact that the o-QFI is maximized at $\phi = n \pi/2$ (for integer $n$) is physically equivalent to having the field-cavity coupling occur at an antinode in a standing wave formed by the beam splitter reflection.  The factor of 2 increase in $\lambda^{(0)}_{\max}$ is therefore equivalent to the factor of 2 increase in a standing wave over its nominal value.  The total contrast in the o-QFI is therefore a factor of 4, consistent with Eq.~(\ref{eq:max_QFI_cs}).

For comparison,  Fig.~\ref{fig:QFI} (b) and (d) shows the numerically computed NOON state o-QFI as a function of $\rho$ and $\phi$ for $N = 2$ and $N = 3$ input photons, respectively.  These plots show that, like the coherent state, the NOON state o-QFI is globally maximized (minimized) at $\rho = 1$ ($\rho = 0$), and therefore also has an enhanced ES sensitivity.  The maximum contrast between min and max o-QFI is $108/64 = 1.6875$, slightly less than that of the coherent state.  Rather than peaking at the antinode, the maximum o-QFI occurs for $\phi = (2n + 1)\pi/4$, suggesting that the NOON state is maximized at a node in the standing wave, where the field has a maximum gradient.  Interestingly, the N = 2 photon NOON state has a smaller maximum QFI than the coherent state with an equal average photon number.  However, for $N \ge 3$ the Heisenberg scaling of the NOON state (QFI $\propto N^2$) overtakes the coherent state's shot-noise limited scaling ($\propto |\boldsymbol{\beta}|^2 = \bar{N}$).

The different values for the maximum contrast in the o-QFI arises simply because $4 \lambda_{\text{abs}}^2$ is generally different from $(\lambda_{\max} - \lambda_{\min})^2$.  When $\rho = 1$ and $\phi$ is an odd multiple of $\pi/4$, Eq.~(\ref{eq:la0_of_z}) simplifies to
\begin{equation}
    \lambda_{+}^{(0)}(\omega) = \frac{2 \gamma\, \omega}{\left(\tfrac{\gamma^2}{4} + \omega^2\right)^2}. 
\end{equation}
Because this is antisymmetric in $\omega$, we can see that for this value of the phase we have $\lambda^{(0)}_{\max} = -\lambda^{(0)}_{\min}$.  We then calculate the extrema by solving $d \lambda_{+}^{(0)}/d\omega = 0$, which yields the value $\lambda^{(0)}_{\max} = \sqrt{27}/\gamma$ with $\omega_{\max} = \gamma/\sqrt{12}$.  
This results in the NOON state QFI having the absolute maximum bound 
\begin{equation}
    \mathcal{I_N} \le 4\, (\lambda^{(0)}_{\max})^2 N^2 = 108\, N^2/\gamma^2
\end{equation}
for the $\epsilon = 0$ ES.

To compare the performance of o-QFI at the corresponding DP, we evaluate Eq.~(\ref{eq:la0_of_z}) for $\rho = 0$ and $\tau = 1$ to find 
$\lambda^{(0)}_{\pm} = \pm \tfrac{\gamma}{\tfrac{\gamma^2}{4} + \omega^2}$.  Interestingly, the eigenvalues reach their respective maxima/minima at the same frequency $\omega_{\max} = \omega_{\min} = 0$, with $\lambda_{\max} = -\lambda_{\min} = 4/\gamma$.  Therefore the NOON state o-QFI at the DP can only reach the value
\begin{equation}
    \mathcal{I_N}^{(dp)} \le (\lambda^{(0)}_{+} - \lambda^{(0)}_-)^2 N^2 = 64 N^2/\gamma^2.
\end{equation}
Thus, the maximum enhancement due to operating on the ES with finite $\rho$ with a NOON state input with $N$ photons is at most $108/64=1.6875$.

We note that at the DP, the maximal value of the NOON state o-QFI, $\mathcal{I_N}^{(dp)} = 64\, N^2/\gamma^2$ is greater than the maximal value of the coherent state QFI, $\mathcal{I}_{\beta}^{(dp)} = 64\,N/\gamma^2$, due to the Heisenberg scaling.  
Finally, we also note that while the optimal NOON state is actually degenerate in frequency, it will instead be an equal quantum superposition over two spatial modes. The first mode is itself a symmetric combination of left and right inputs, while the second is an antisymmetric combination.   

\subsection{Properties of the local generator $\mathbf{A}_\epsilon$}
\begin{figure*}[hbtp]
    \includegraphics[width=2.0\columnwidth]{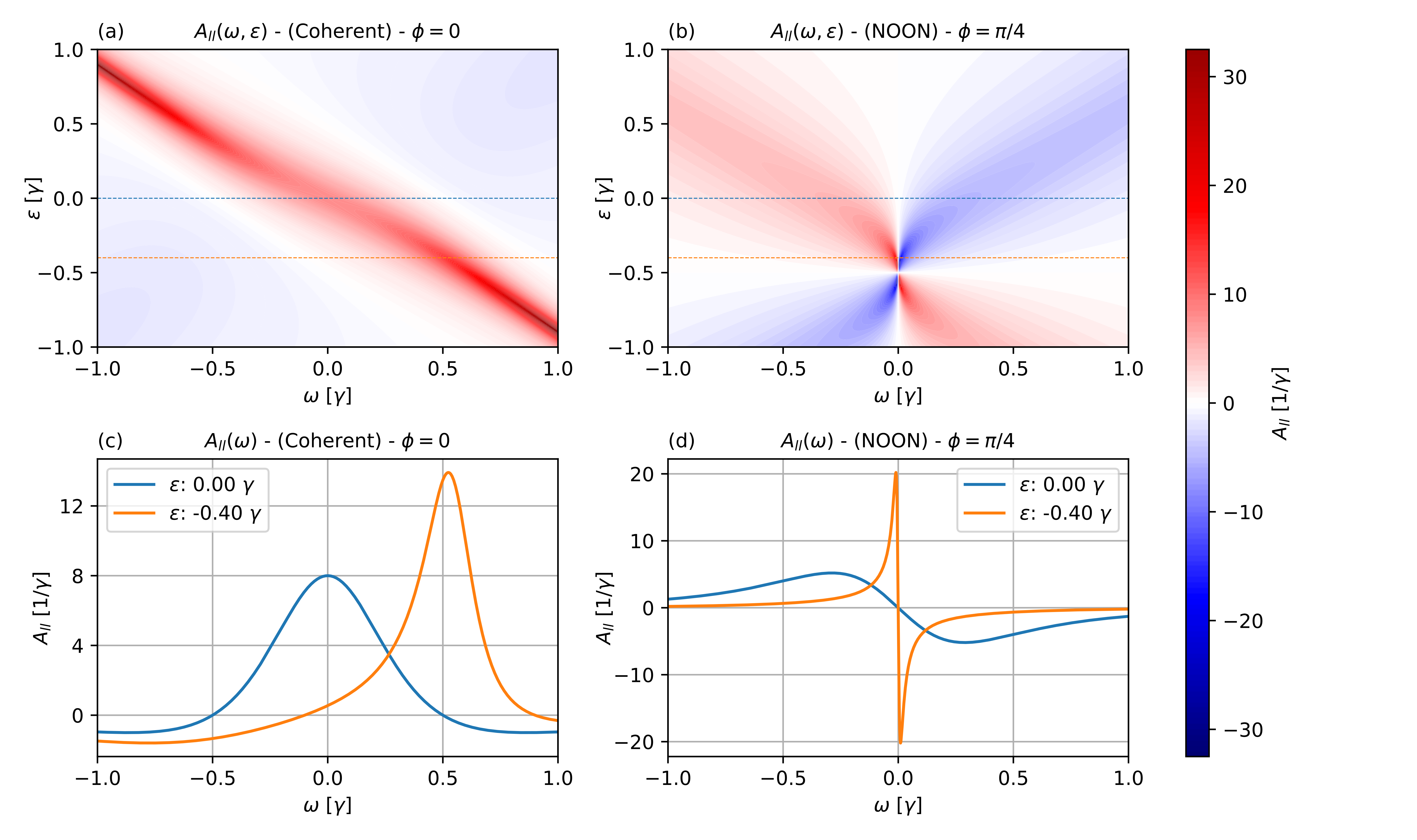}
    \caption{\label{fig:A_eps} 
   [color online] Panels (a) and (b) show contour plots of $A_{ll}(\omega, \epsilon)$ for complete reflection ($\rho = 1$), when $A_{ll}$ is the only non-zero matrix element of $\mathbf{A}_{\epsilon}$.  In panel (a) the phase is fixed for the coherent state optimal value of $\phi = 0$, and in panel (b) the phase is fixed at the NOON state optimal value of $\phi = \pi/4$ (see Fig.~\ref{fig:QFI}).  Panels (c) and (d) show $A_{ll}(\omega)$ for constant values of $\epsilon$ with $\phi = 0$ in (c) and $\phi = \pi/4$ in (d). The coupling parameter values are for the ES $\epsilon = 0$ (blue lines) and the non-degenerate value of $\epsilon = -0.4 \gamma$ (orange lines).  When $\phi = \pi/4$, $A_{ll}(\omega, \epsilon)$ becomes ill-defined at $\omega = 0$ and $\epsilon = -0.5\, \gamma$, showing sharply peaked oscillations in panel (d). At these values, $\tilde{ \mathbf{H} }_{\epsilon}$ has the eigenvalues $\{ -i\gamma, 0 \}$, and so the resolvent $\mathbf{R}_{\epsilon}(\omega)$ is undefined at $\omega = 0$.} 
\end{figure*}

We have shown above that as $\rho$ increases from 0 to 1, the optimal $\epsilon = 0$ QFI also increases from a minimal value on the DP ($\rho = 0$), to a maximal value on the ES at the location $\rho = 1$ and with a phase $\phi$ that depends on the choice of input state.  However, this does not indicate whether this improvement is directly caused by the presence of an EP in the system.  If the root cause of this increase is due to the ES, then the surface defined by $\epsilon = 0$ should contain the largest obtainable QFI, even when maximized over \emph{all} possible parameter values including $\epsilon$ itself.  However, here we demonstrate that for the asymmetric microring resonator model of Figure~\ref{fig:schematic}, this is not the case and that for both coherent and NOON states, the optimal value of the QFI can actually increase for values of $\epsilon$ trending away from the ES.  This implies that for sensing non-perturbatively small values of $\epsilon$, greater sensitivity could actually be achieved by operating at points not on the ES. 

To understand this phenomenon, we note first that on the ES the QFI is maximized for $\rho = 1$.  Therefore, we examine the spectrum of $\mathbf{A}_\epsilon$ for the fixed value of $\rho = 1$.   From Eq.~(\ref{eq:A_ij}), it is evident that when $\rho = 1$ (and thus $\tau = 0$), the only non-zero matrix element is $A_{ll}$. 
Therefore, the two eigenvalues of $\mathbf{A}_{\epsilon}$ are $A_{ll}$ and 0, i.e., if $A_{ll} > 0$ ($A_{ll} < 0$) then $\lambda^{(\epsilon)}_{+} = A_{ll}$ ($\lambda^{(\epsilon)}_{-} = A_{ll}$), respectively, and the remaining eigenvalue is zero.  
Figure~\ref{fig:A_eps} shows the $\rho=1$ behavior of $A_{ll}$ for the values of $\phi$ where the coherent state, panels (a) and (c), and NOON state, panels (b) and (d), obtained the largest o-QFI.  
Specifically, panels (a) and (b) show $A_{ll}$ as a contour plot over the two remaining free parameters $\omega$ and $\epsilon$. The lower panels (c) and (d) show horizontal slices at fixed $\epsilon = 0$ (blue lines) and $\epsilon = -0.4\, \gamma$ (orange lines). The $\epsilon = 0$ slice in panel (c) reveals that the value of $\omega = 0$ maximizes the QFI for the coherent state, while panel (d) shows that for the NOON state the QFI eigenvalues $\lambda_{\max}$ ($\lambda_{\min}$) occur at $\omega = \gamma/ \sqrt{12}$ ($\omega = -\gamma/\sqrt{12}$), respectively.  Note that for $\epsilon = 0$ (the blue lines) the absolute maximum in panel (c) is larger than the maximum in panel (d), which shows why, for the coherent state input, the phase $\phi = 0$ is optimal.  In contrast, it turns out that when $\phi = 0$, the maximum eigenvalue difference $\lambda_{\max} - \lambda_{\min} = 9/\gamma$, while when $\phi = \pi/4$, $\lambda_{\max} - \lambda_{\min} = 2 \sqrt{27}/\gamma \approx 10.4 \gamma$, implying that the phase $\phi = \pi/4$ gives a larger QFI for a NOON state input.  

However, close analysis of panels (a) and (b) shows that these peaks on the $\epsilon = 0$ surfaces are not global optima.  When $\phi = 0$, (panel (a)) the origin $(\omega=0, \epsilon=0)$ constitutes a saddle point with increasing gradient on the line $\epsilon \approx -0.68\, \omega$ and decreasing gradient along the line perpendicular to this. (It is interesting to note that, along the true diagonal $\epsilon = -\omega$, $A_{ll}(\epsilon, \omega)$ has a constant value of $8/\gamma$.)  

A more complex scenario occurs for the NOON state input with $\phi = \pi/4$.  For this case, Figure~\ref{fig:A_eps} (b) shows an increasingly large variation of $A_{ll}(\omega, \epsilon)$ until an apparent divergence occurs at the point $(\omega = 0, \epsilon = -\gamma/2)$.  This divergence stems from a resonance in 
$\tilde{\mathbf{H}}_{\epsilon}$ at $\epsilon = - \gamma/2$, such that at these parameter values the two Hamiltonian eigenvalues become $\{0, -i \gamma\}$.  The existence of a zero eigenvalue means that $\operatorname{det}(\tilde{\mathbf{H}} - \omega\cdot \mathbf{I} )|_{\omega = 0} = 0$ and thus the resolvent $\mathbf{R}_{\epsilon}(0)$ is undefined at $\epsilon = - \gamma/2$.  Figure~\ref{fig:A_eps}(d) further reveals that along the orange line $\epsilon = -0.4 \gamma$,  $A_{ll}(\omega, \epsilon)$ has a larger variation than that seen along the blue $\epsilon =0$ line. This behavior increases on further approaching the singularity in panel (d) as $\epsilon \rightarrow -\gamma/2$, apparently without bound.   The physical effect of this resonance is that the phase of $K_{ll}(\omega)$ becomes equal to $\pi/2 + 2 \gamma \omega (\epsilon + \gamma/2)^{-2} + O(\omega^3)$ which is singular when $\epsilon \rightarrow -\gamma/2$. (See Sec.~\ref{sec:NOON_measurement} and Eq.~(\ref{eq:K_ll_NOON}) below.)
In practice, any unavoidable optical losses which are ignored here will broaden this resonance and place a finite upper bound on $A_{ll}(\omega, \epsilon)$ at this point.

\section{Detection schemes\label{sec:detection}}

While the QCRB shows what sensitivity is in principle achievable, it does not readily indicate what practical measurement saturates this bound.  Given that the QFI is the maximum classical Fisher information generated by any POVM, any specific measurement can be compared to the QFI by computing the classical Fisher information for the observed probabilities.  Thus given a specific measurement, if one shows that the classical Fisher information agrees with the QFI, then that measurement is optimal.   Here we show that for a coherent state input, a homodyne detection scheme achieves the QCRB, and that for a NOON state input, a different measurement, namely a frequency-shifted photon-counting interferometer, saturates the QCRB.  Thus in both cases it is possible to undertake optimal sensing of a perturbatively small cavity mode coupling parameter $\epsilon$.

\subsection{Coherent state - Homodyne detection}

\begin{figure}[htbp]
    \includegraphics[width=1.0\columnwidth]{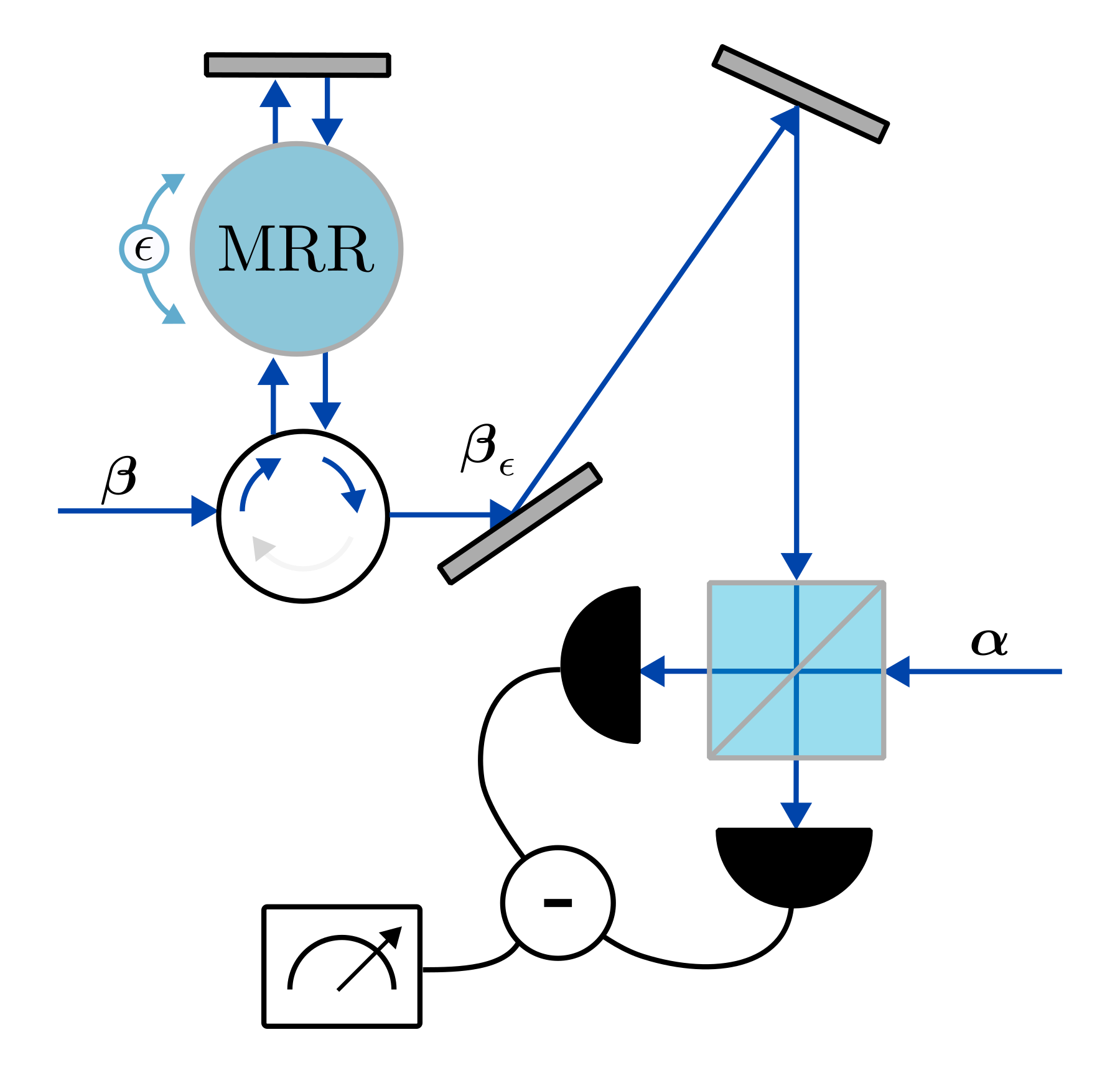}
    \caption{\label{fig:homodyne-measurement}[color online] Balanced homodyne detection.  The input coherent state $\boldsymbol{\beta}$, interacts with the system and is spatially separated from the input by a circulator.  The $\epsilon$ dependent output $\boldsymbol{\beta}_\epsilon$, is then interfered on a symmetric 50:50 beam splitter with a large amplitude coherent state $\boldsymbol{\alpha}$.  The difference in photon counts gives a mean signal $\mu = 2 \operatorname{Im} \langle \boldsymbol{\alpha}, \boldsymbol{\beta}_\epsilon \rangle$ and a variance $\sigma^2 = \|\boldsymbol{\alpha}\|^2 + \|\boldsymbol{\beta}\|^2$.  When $\|\boldsymbol{\beta}\| \ll \|\boldsymbol{\alpha}\|$, the difference signal is a normal random variable of mean $\mu$ and variance $\sigma^2$.}
\end{figure}

For a normal random variable whose mean $\mu(\epsilon)$ depends upon $\epsilon$, while its variance, $\sigma^2$ does not, the classical Fisher information is~\cite{lehmann_theory_1998}
\begin{equation}\label{eq:I_norm}
    \mathcal{I}_{norm} = \left(\frac{\partial \mu}{\partial \epsilon} \right)^2 \frac{1}{\sigma^2}. 
\end{equation}
We pause to note that Eq.~(\ref{eq:I_norm}) emphasizes that when maximizing the sensitivity of a measurement to $\epsilon$, one does not want to maximize the direct signal-to-noise ratio, i.e., $\mu/\sigma$.  Rather, one wishes to maximize the change in the mean of the measurement relative to the noise, i.e., $(\frac{d \mu}{d \epsilon}/\sigma)^2$.
For a given input coherent state (with a vector amplitude $\boldsymbol{\beta}(\omega)$), balanced homodyne detection computes the difference in photon counts between the output coherent amplitude $\boldsymbol{\beta}_\epsilon$ and an input coherent state $\boldsymbol{\alpha}$, as illustrated in  Figure~\ref{fig:homodyne-measurement}. The mean $\mu$ and variance $\sigma^2$ of the photon difference are easily calculated,~\cite{scully_quantum_optics_1997} yielding $\mu = 2 \operatorname{Im} \langle \boldsymbol{\alpha}, \boldsymbol{\beta}_\epsilon \rangle$ and $\sigma^2 = \|\boldsymbol{\alpha}\|^2 + \|\boldsymbol{\beta}\|^2$.  
Utilizing Eq.~(\ref{eq:beta_eps}) and Eq.~(\ref{eq:A_as_GWSM}), we have
\begin{equation}
\begin{split}
    \frac{d\mu}{d\epsilon} &= 2 \operatorname{Im} \langle \boldsymbol{\alpha}, \frac{d}{d\epsilon} \mathbf{K}_\epsilon \boldsymbol{\beta} \rangle\\
     &= 2 \operatorname{Im} \langle \boldsymbol{\alpha}, i \mathbf{K}_\epsilon \mathbf{A}_\epsilon \boldsymbol{\beta} \rangle\\
     &= 2 \operatorname{Re} \langle \mathbf{K}_\epsilon^\dag\boldsymbol{\alpha},\,  \mathbf{A}_\epsilon \boldsymbol{\beta} \rangle.
\end{split}
\end{equation}
When $\boldsymbol{\beta}$ is an eigenvector of $\mathbf{A}_\epsilon$, $\boldsymbol{\beta}_{abs}$, with eigenvalue $\lambda_{abs}$ we then have
\begin{equation}
\begin{split}
    \left|\frac{d\mu}{d\epsilon}\right| &= 2\, | \lambda_{abs}\, \operatorname{Re} \langle \mathbf{K}_\epsilon^\dag \boldsymbol{\alpha},\, \boldsymbol{\beta}_{abs} \rangle | \\
    &\le  2\, |\lambda_{abs} | \, \| \boldsymbol{\alpha}\|\, \|\boldsymbol{\beta}_{abs} \|.
\end{split}
\end{equation}
The second line derives from an application of the Cauchy-Schwartz inequality, which is saturated when $\alpha \propto \boldsymbol{K}_\epsilon \boldsymbol{\beta}_{abs}$. In other words if
\begin{equation}
    \boldsymbol{\alpha}_{opt} = \sqrt{N_\ell}\, \mathbf{K}_\epsilon \boldsymbol{\beta}_{abs} / \|\boldsymbol{\beta}_{abs}\|
\end{equation}
then 
\begin{equation}
\begin{split}
    |\operatorname{Re} \langle \mathbf{K}_\epsilon^\dag \boldsymbol{\alpha}_{opt},\, \boldsymbol{\beta}_{abs} \rangle| &= \sqrt{N_\ell}\, | \operatorname{Re} \langle \mathbf{K}_\epsilon^\dag \mathbf{K}_\epsilon \boldsymbol{\beta}_{abs},\, \boldsymbol{\beta}_{abs} \rangle| / \|\boldsymbol{\beta}_{abs}\| \\
    &= \sqrt{N_\ell}\, | \operatorname{Re} \langle \boldsymbol{\beta}_{abs},\, \boldsymbol{\beta}_{abs} \rangle| / \|\boldsymbol{\beta}_{abs}\|\\
    &= \sqrt{N_\ell}\, \|\boldsymbol{\beta}_{abs}\|,
\end{split}
\end{equation}
which indeed saturates the above Cauchy-Schwartz inequality. Here we set the average local oscillator photon number to be $\|\boldsymbol{\alpha}_{opt}\|^2 \equiv N_{\ell}$, where for homodyne detection $N_{\ell} \gg \|\boldsymbol{\beta}_{abs}\|^2$. 

Thus the optimal Fisher information is
\begin{equation}
\begin{split}
    \mathcal{I}_{norm} &= (2\, \lambda_{abs}\,  \| \boldsymbol{\alpha}_{opt}\|\, \|\boldsymbol{\beta}_{abs} \| )^2 / \sigma^2\\
    &= 4\,\lambda_{abs}^2\, \|\boldsymbol{\beta}_{abs} \|^2\,\frac{N_{\ell} }{ N_{\ell} + \|\boldsymbol{\beta}_{abs} \|^2} \\
    &= 4\,\lambda_{abs}^2\, \|\boldsymbol{\beta}_{abs} \|^2 + O( \tfrac{ \|\boldsymbol{\beta}_{abs} \|^2}{N_\ell} ),
    \end{split}
\end{equation}
which is equal to the QFI for coherent input states in Eq.~(\ref{eq:QFI_cs}) when $\tfrac{\|\boldsymbol{\beta}_{abs} \|^2}{N_\ell} \rightarrow 0$. 

As a final note, we have also compared the above coherent state QFI to the signal-to-noise ratio (SNR) computed in Ref.~\onlinecite{lau_fundamental_2018}.  The setting of that work is to consider a coherent state introduced via a single input port, which then couples to a linear network of oscillators.  This network is subjected to a perturbation proportional to $\epsilon \ll 1$.  The authors considered the statistics of a homodyne measurement of the reflected input and calculated a limiting bound on the SNR.  In Appendix~\ref{app:Lau_clerk} we translate our system to the framework of Ref.~\onlinecite{lau_fundamental_2018} to relate the maximum SNR to the QFI, showing that the maximum SNR is precisely equal to $\epsilon^2\, \mathcal{I}_\beta (0)$.  

\subsection{NOON state - frequency shifted photon counting}\label{sec:NOON_measurement}

\begin{figure}[htbp]
    \includegraphics[width=1.0\columnwidth]{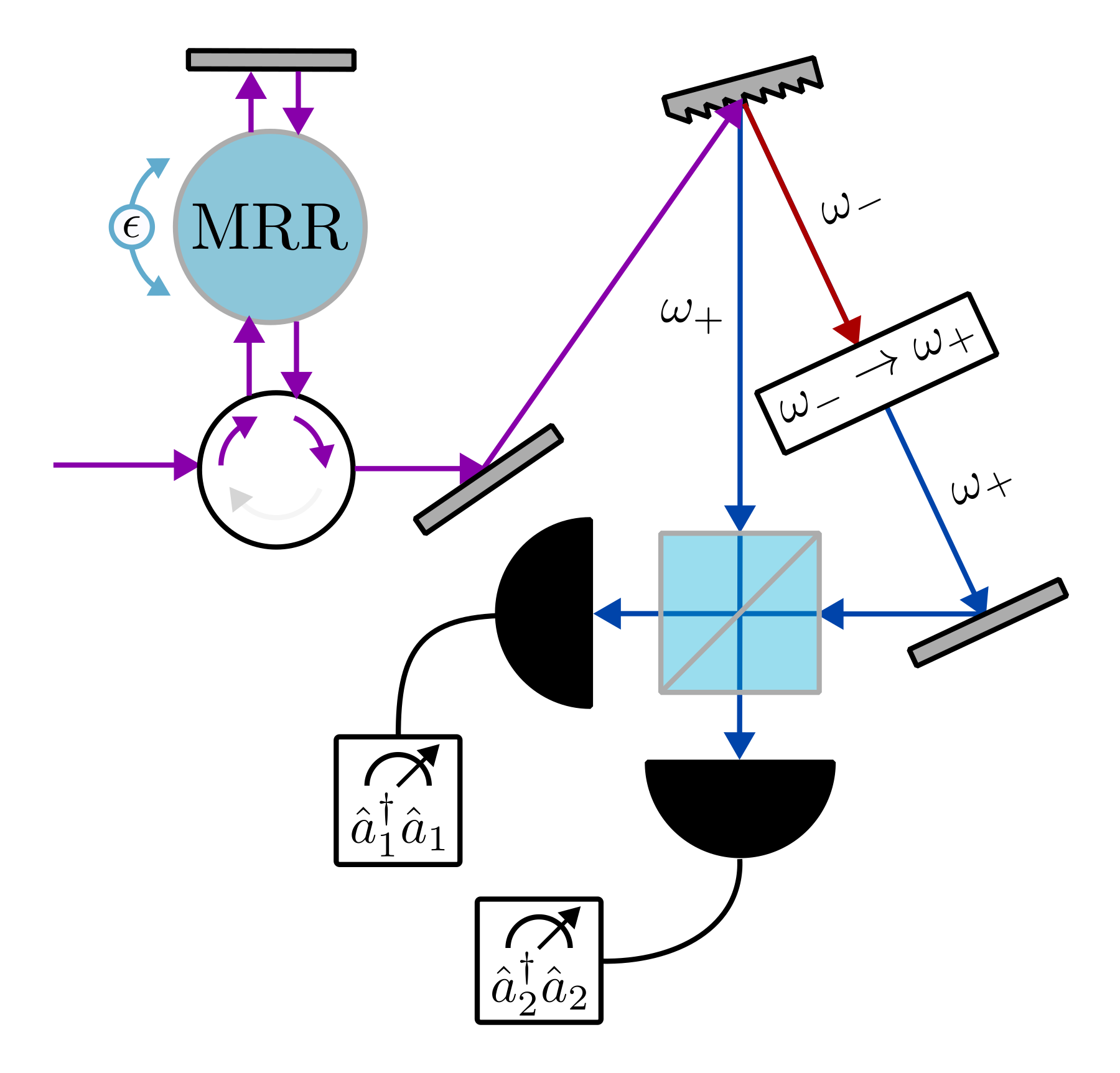}
    \caption{\label{fig:NOON-measurement}[color online] NOON state detection scheme.  The reflected NOON state is spatially separated from the input by a circulator.  The frequency content of the output is dispersed into two distinct spatial modes.  The lower frequency $\omega_-$ is shifted to match $\omega_+$ via a high efficiency optical modulator.  The common frequency beams interfere on a 50:50 beam splitter.  The mixed outputs are measured at photon counters 1 and 2.}
\end{figure}

At the NOON state optimum values of $\rho = 1$ and $\phi = \pi/4$,
\begin{equation}
    \mathbf{K}_{\epsilon} = \left[ \begin{matrix} K_{ll}(\omega) & 0 \\ 0 & -1 \end{matrix} \right]
\end{equation}
with $K_{ll}(\omega)$ from Eq.~(\ref{eq:Kll}) simplifying to
\begin{equation} \label{eq:K_ll_NOON}
    K_{ll}(\omega) = i\frac{(\omega - i \gamma/2)^2 - \epsilon(\epsilon + \gamma)}{(\omega + i \gamma/2)^2 - \epsilon(\epsilon + \gamma)} \equiv e^{i(2 \chi(\omega, \epsilon) + \pi/2)}.
\end{equation}
We define the phase $\chi(\omega, \epsilon) \equiv \operatorname{arg}[ (\omega - i \gamma/2)^2 - \epsilon(\epsilon + \gamma)]$. 
Factoring out the irrelevant global phase of $\pi/2$ allows us to utilize the symmetry $\chi(-\omega, \epsilon) = -\chi(\omega, \epsilon)$.  In Appendix~\ref{app:NOON}, we represent a NOON state by the $N$-particle wave function $|\textrm{NOON}\rangle = \frac{1}{\sqrt{2}} \left( \psi_1^{\otimes N} + \psi_2^{\otimes N}\right) $.  When $\psi_i$ is a monochromatic plane wave with frequency $\omega_i$ interacting with the cavity, we have
\begin{equation}
    \mathbf{K} |\omega_i\rangle^{\otimes N} = i e^{2 N \chi(\omega_i) } |\omega_i\rangle^{\otimes N}.  
\end{equation}
We now choose the frequencies $\omega_{i}$ so that we obtain the o-QFI at a chosen nominal value of $\epsilon$, e.g., $\epsilon = 0$.  In other words, we fix $\omega_1 = \omega_{\max}$ and $\omega_2 = \omega_{\min}$.  When $\epsilon = 0$, $\omega_{\max} = - \omega_{\min} = \gamma/\sqrt{12}$.  We define the phase $\theta(\epsilon) = 2 \chi(\omega_{\max}, \epsilon)$, so that the reflected output state is 
    \begin{equation}
        |MRR_{out}\rangle = \frac{i e^{i N \theta(\epsilon) }}{\sqrt{2}} |\omega_+\rangle^{\otimes N} + \frac{i e^{-i N \theta(\epsilon) }}{\sqrt{2}}|\omega_-\rangle^{\otimes N}
    \end{equation}
This output state is nearly the relative phases induced in the two arms of an interferometer.  The only difference is that when interfered on a beam splitter the states $|\omega_+\rangle$ and $|\omega_-\rangle$ are orthogonal and therefore will fail to produce the required interference.  

To alleviate this problem, we consider the idealized detection setting depicted in Figure~\ref{fig:NOON-measurement}.  The reflected light exiting the microring resonator is directed to a dispersive element (e.g., a diffraction grating), so that the bichromatic NOON state becomes spatially separated.  One of the two output beams is then frequency shifted by a high-efficiency optical modulator, so that the NOON state is transformed from a superposition that is entangled between spatial mode and frequency to a superposition that is degenerate in frequency.  When mixed on a 50:50 beam splitter the output state then shows interference between the two paths, i.e.
\begin{multline}
    |\psi_{out}\rangle = i \cos\big[ N \theta(\epsilon)\big]\,  |\omega_+\rangle^{\otimes N}\otimes|0\rangle  \\
    -  \sin\big[ N \theta(\epsilon)\big]\, |0\rangle\otimes|\omega_+\rangle^{\otimes N},
\end{multline}
where the tensor product state $|\omega_+\rangle^{\otimes N}\otimes |0\rangle$ represents $N$ photons with frequency $\omega_+$ at the first output and zero photons at the second output. 

This NOON state then has only two photon-counting outcomes with nonzero probability, namely, either $N$ photons are measured at detector 1 and none at detector 2, or vice versa. These two outcomes have the probabilities:
\begin{equation}
    \begin{split}
        P_{1} &= \cos^2( N \theta\,) \text{ and}\\
        P_{2} &= \sin^2( N \theta\,)
    \end{split} 
\end{equation}
respectively. The fact that all photons are measured by detector 1 or by detector 2 with no cross-detector coincidences shows that this detection scheme is an $N$-photon version of a Hong-Ou-Mandel (HOM) inteferometer with maximum contrast~\cite{loudon_quantum_2000}.

To show that this measurement does achieve the QRCB, we calculate the classical Fisher information, which for this two-outcome measurement has the simple form
\begin{equation}
\begin{split}
    \mathcal{I}_{NOON} &= \frac{1}{P_1}\left ( \frac{\partial P_1 }{\partial \epsilon} \right)^2 + \frac{1}{P_2}\left ( \frac{\partial P_2 }{\partial \epsilon} \right)^2 \\
    &= 4 N^2\, \left ( \frac{d \theta(\epsilon) }{d \epsilon} \right)^2.
\end{split}
\end{equation}
At the parameter values $\rho =1$ and $\phi = \pi/4$,  we have the relation 
\begin{equation}
    \begin{split}
    \frac{\partial K_{ll}(\omega_{\max})}{\partial \epsilon} &= \frac{d}{d \epsilon} e^{i \theta(\epsilon) + i \pi/2 } =  + i K_{ll}(\omega_{\max}) \frac{d \theta(\epsilon)}{d \epsilon}\\
    &= i K_{ll}(\omega_{\max}) A_{ll}(\omega_{\max}) 
    \end{split}
\end{equation}
where the second line followed from Eq. (\ref{eq:A_as_GWSM}). 
Thus we infer that $\frac{d \theta}{d\epsilon} = A_{ll}(\omega_{\max})$.  However, by definition $\omega_{\max}$ maximizes $A_{ll}(\omega)$ at the chosen value of $\epsilon$, e.g. $\epsilon = 0$.  It can be seen in Fig.~\ref{fig:A_eps}(b) that for this choice in phase $A_{ll}(\omega) = - A_{ll}(-\omega)$, which results in the relation
\begin{equation}
    ( \lambda_{max} - \lambda_{min} )^2 = ( A_{ll}(\omega_{max}) - A_{ll} (\omega_{\min}) )^2 = 4 \left( \frac{d \theta}{d \epsilon}\right)^2.
\end{equation}  
This shows that
$\mathcal{I}_{NOON} = N^2\, (\lambda_{max} - \lambda_{min})^2 = \mathcal{I_N}(0)$ saturating the QCRB.

\section{discussion and conclusions\label{sec:conclusion}}

In this work we have studied a simple model of a bidirectional cavity interacting with a retro-reflected output from a partially reflecting mirror with reflectivity $\rho$.  This produces a non-Hermitian Hamiltonian, which is exceptional when $\rho > 0$ and the cross coupling between modes $\epsilon$ is zero. The Hamiltonian remains exceptional as the remaining parameters are varied, thus defining an exceptional surface.  The quantum Cram\'er-Rao bound states that an unbiased estimator of $\epsilon$ has a mean-squared error that is bounded from below by the inverse of the quantum Fisher information (QFI) evaluated at the unknown value of $\epsilon$.

The fully quantum analysis carried out here shows that experimental parameters can be chosen so that the sensitivity of the exceptional points ($\rho > 0$) are greater than corresponding diabatic points at $\rho = 0$.  The optimal parameters and final QFI critically depends upon the phase $2\phi$ acquired when reflecting off the mirror.  In particular, when the incident light is in a coherent state, the QFI is maximized at a phase where the reflected output forms interference pattern that places an antinode at the location of the cavity, while no enhancement is found for a nodal configuration.  In contrast, a nonclassical NOON state has an optimal phase shift that places the cavity at a point of maximal slope in the standing wave.  Analysis of the behavior over the entire range of reflectivity between $\rho = 0$ and $\rho = 1$, shows that the QFI exhibits an enhancement that is at most a factor of 4 for a coherent input state, while for a NOON input state with $N$ photons, the enhancement is at most a factor of 1.6875. The analysis also showed that the NOON state exhibits Heisenberg scaling with photon number $N$, resulting in greater values of the QFI for $N \geq 3$.  In addition to analyzing the QFI, we have identified experimental configurations for both coherent and NOON state inputs that allow saturation of the quantum Cram\'er-Rao bound.

Our analysis revealed that the key factor determining the magnitude of the QFI is the spectrum of the Hermitian operator $\mathbf{A}_\epsilon$, which is the generator for local changes in $\epsilon$. The absolute magnitude of the maximum eigenvalue of $\mathbf{A}_\epsilon$ determines the optimal QFI for coherent state inputs, while the spectral range of $\mathbf{A}_\epsilon$ determines the optimal value for NOON state inputs.  An interesting corollary of our analysis is then that when considering the sensitivity on an absolute scale, it is not possible to conclude that the presence of an ES in the current asymmetric system is the deciding factor in optimal sensing of arbitrary values of the $\epsilon$ parameter.  This is because the eigenvalues of $\mathbf{A}_\epsilon$ are not generally peaked around points on the exceptional surface located at $\epsilon = 0$.  Specifically, near the parameter values $(\rho, \phi) = (1,0)$ the nonzero eigenvalue of $\mathbf{A}_0$ forms a saddle point at $(\omega, \epsilon) = (0,0)$, see Figure~\ref{fig:A_eps} (b).  Presumably if the exceptional nature at this point on the ES were a key factor then $(\omega, \epsilon) = (0,0)$ would be at least a local maximum that decays with increasing $|\epsilon|$.   

The fact that the QFI is directly related to the spectrum of $\mathbf{A}_{\epsilon}$, shows that when maximizing a system's sensitivity, the natural route is then to improve the maximum bounds of the spectrum.  Here we determined that the eigenvalues of $\mathbf{A}_{\epsilon}$ diverge near points where $\tilde{\mathbf{H}}$ has strictly real eigenvalues.  This suggests that to optimize the sensing of arbitrary values of $\epsilon$ it would be best to investigate the QFI in a $\mathcal{PT}$-symmetric system where the spectrum of $\tilde{\mathbf{H}}_\epsilon$ is entirely real.  Creating such a system requires introducing amplifying gain, which in turn introduces new sources of noise.  Previous models have  included gain via an input source of squeezed light~\cite{lau_fundamental_2018}, which introduced gain via a modified version of our $\mathbf{B}$ matrix.  However, this does not change the spectrum of $\tilde{\mathbf{H}}_\epsilon$ and therefore provides only a limited benefit.  Including a truly $\mathcal{PT}$-symmetric system Hamiltonian would require studying a system like an optical parametric amplifier ~\cite{roy_nondissipative_2021}.

\begin{figure}[htbp]
    \includegraphics[width=1.0\columnwidth]{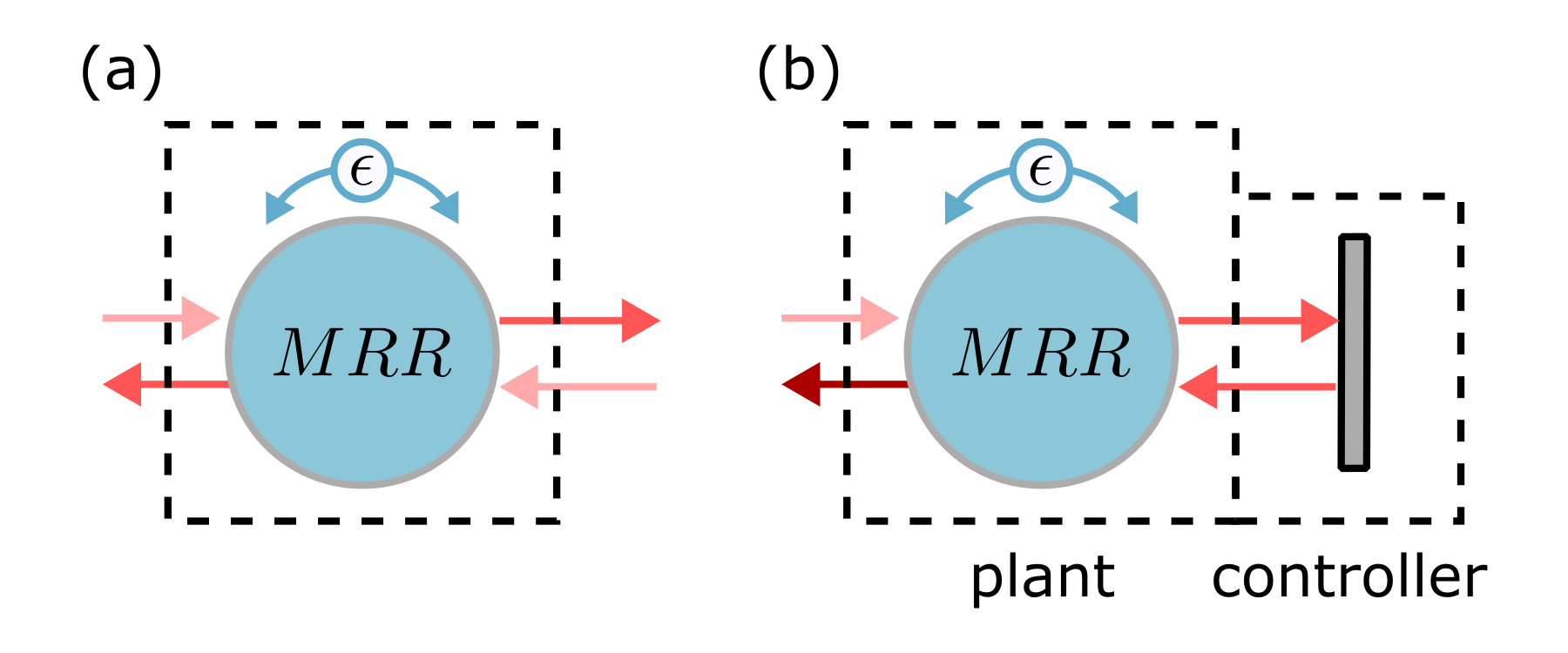}
    \caption{\label{fig:qucontrol}[color online] Quantum control schematics. (a) When $\rho = 0$ we have an open control setting, where both the left and right inputs are externally applied, without reference to the unknown $\epsilon$. (b) When $\rho = 1$ we have a closed loop coherent feedback setting in which the resonator is the ``plant'' to which coherent proportional feedback is applied by the external mirror, which acts as the ``controller''.}
\end{figure}

One possible explanation for the sensing improvements seen here on the ES is that the retro-reflection acts as a source of coherent feedback. Figure~\ref{fig:qucontrol} (a) depicts the pure transmission $\rho = 0$ setting, with no retroflection, where there are two free inputs and two free outputs.  When computing the maximum QFI in this case,  we optimize both the left and right input fields in order to maximize the total information contained in both output fields.  Neither of these inputs contains any dependence upon the unknown value of $\epsilon$ and thus can only be optimized for a given nominal value of $\epsilon$.  

This lack of information in the pure transmission scenario $\rho=0$ is compared with the $\rho = 1$ scenario of total reflection, which is shown in panel (b) of Figure~\ref{fig:qucontrol}.  Here, only the left-side input is under the control of the experimentalists and is optimized for a nominal value of $\epsilon$. However, here the entirety of the right-side input is the reflected output from the clockwise (CW) cavity mode.  The key point here is that this configuration can yields more information about $\epsilon$, since the right-side (retro-reflected) input will now contain some information about the unknown parameter $\epsilon$ that was gained in the first pass from left to right.  If the reflection phase shift $\phi$ is optimally chosen, then the second pass output contains more information than the combined outputs in the $\rho = 0$ configuration.

This retro-reflection scheme performs a simple form of coherent feedback control~\cite{gough_principles_2012}.  As depicted in Figure~\ref{fig:qucontrol} (b), the microring resonator acts as a targeted quantum system, often called the ``plant'' in the classical control literature. The mirror acts as a coherent ``controller'' that provides a particularly simple form of proportional feedback control.   With the correct tuning, the final output signal can then show an increased sensitivity to the unknown parameter $\epsilon$.

Finally, we note that by appealing to the quantum Cram\'er-Rao bound, we have limited our discussion here to unbiased point estimates of $\epsilon$.  A useful avenue of future research would be to consider more sophisticated setting such as a Bayesian estimator for a given prior and/or procedures that are robust to likely sources of error e.g. photon loss or finite pulse bandwidths.

\begin{acknowledgments}
We thank Qi Zhong, Ali Kecebas, Sahin K. Özdemir, and Stefan Rotter for helpful discussions and insights.  
This work was supported by the AFOSR Multidisciplinary University Research Initiative (MURI) Award on Programmable systems with non-Hermitian quantum dynamics (Award No. FA9550-21-1-0202). 
Additional support is acknowledged from the U.S. Department of Energy, Office of Science, National Quantum Information Science Research Centers, Quantum Systems Accelerator.

\end{acknowledgments}

\section*{Data Availability Statement}

The data and code that support the findings of this article are available under an open source license~\cite{Cook_optimal-ES-QFI}.

\appendix

\section{Input-Output derivation of the Heisenberg equations of motion}\label{app:IO_HLE}
Here we derive the Heisenberg equations of motion for the cavity field, using the input-output formalism of quantum optics~\cite{gardiner_quantum_2004,jacobs_quantum_measurement_theory_2014,cook_input-output_2018}.  
In such a theory, a quantum system (our two cavity modes) are resonantly coupled to a countable number of one-dimensional input field modes, $\hat{a}^{in}_{i}(t)$.  These input fields are taken to be slowly varying on a time scale that is fast when compared to the cavity resonance frequency, $\omega_{cav}$.  The field operators are ``slowly-varying'' in the sense that they represent the input changes and fluctuations of the free electromagnetic field over a bandwidth $\Delta \omega \ll \omega_0$.  Under a Markov approximation the field operators become delta-commuting quantum white-noise, such that for times $t, t' \gg 1/\omega_0$, $[\hat{a}^{in}_{i}(t), \hat{a}^{in\,\dagger}_{j}(t')] = \delta_{ij}\, \delta( t - t')$. 

The input-output formalism propagates both how the local quantum system responds to the input fields and how the local system changes the output fields.  The entire interaction is specified by the triple $(\mathbf{S}, \mathbf{L}, H)$~\cite{combes_slh_framework_modeling_2017}.  The unitary matrix $\mathbf{S}$ controls how the input field operators $\hat{a}^{in}_j$ are scattered to the output modes $\hat{a}^{out}_i$.  In general, the elements $S_{ij}$ could depend upon system operators, but here we assume that they are simple complex numbers, e.g., the entries of the transmission and reflection coefficients of a unitary beam-splitter.  The components of the vector $\mathbf{L}$, $L_i$,  are system operators that indicate how the local system acts as a source for output mode $i$.  Finally, $H$ is the local Hamiltonian, which generates the system's coherent evolution.  

When $\mathbf{S}$ is independent of the local system (i.e., it commutes with any system observable), the time evolution of any system operator $X$ is given by the Heisenberg-Langevin equation (HLE)~\cite{cook_input-output_2018}
\begin{equation}
\label{eq:HLeqs}
\begin{split}
    \dot{X} &= i\left[H, X \right]  - \tfrac{1}{2} \left(  [X, \mathbf{L}^\dag] \mathbf{L} - \mathbf{L}^\dag [X,\mathbf{L}]  \right) \\
  &\quad - [X,\mathbf{L}^\dag ] \mathbf{S} \hat{\mathbf{a}}^{in} + \hat{\mathbf{a}}^{in\,\dagger}\mathbf{S}^\dagger [X, \mathbf{L} ].
\end{split}
\end{equation}
For simplicity, we choose to work in units where $\hbar = 1$.  Since this expression is in the Heisenberg picture, the operators $X$, $\hat{\mathbf{a}}^{in}$, $\mathbf{L}$, etc. are time-dependent, which we have notationally suppressed for clarity. We note that if the input fields are in the vacuum state, averaging over the state of the field results in any terms proportional to $\hat{\mathbf{a}}^{in}$ and $\hat{\mathbf{a}}^{in\,\dagger}$ averaging to zero, reducing this HLE to the Heisenberg form of a Lindblad master equation with jump operators $\hat{L}_i$.  Thus in absence of the BS, the cavity modes decay with the rate $\gamma$.   This corresponds to having $\mathbf{L}^{cav} = \sqrt{\gamma} \hat{\mathbf{a}}^{cav}$. However, in the presence of the BS, $\mathbf{L}$ takes on a different form. 

The vector of outputs, $\hat{\mathbf{a}}^{out}(t)$, is not generally independent from the local system due to the coupling of this to the outputs expressed via $\mathbf{L}$. In fact, $\mathbf{L}$ plays two roles; the first is in the above HLE for system operators $X$, while the second is its action as a source for the output fields.  
Specifically, the output fields $\hat{\mathbf{a}}^{out}(t)$ are related to the inputs and the system via the simple equation~\cite{combes_slh_framework_modeling_2017}
\begin{equation}\label{eq:IO_relation_general}
    \hat{\mathbf{a}}^{out}(t) = \mathbf{S}\, \hat{\mathbf{a}}^{in}(t) + \mathbf{L}(t).
\end{equation}
We note that because $\hat{\mathbf{a}}^{out}(t)$ has units of $1/\sqrt{\text{s}}$,  $\mathbf{L}$ will have the same units, hence the $\sqrt{\gamma}$ factor in $\mathbf{L}^{cav}$.   Also, a change in phase in the components of $\mathbf{L}$ changes the phase relationship between the scattered inputs and the source fields from the system.  So if any additional phase shift are needed, the entries of $\mathbf{S}$ and $\mathbf{L}$ can be easily modified to reflect this.  The only constraint is that $\mathbf{S}$ must be a unitary matrix.  

When computing the Heisenberg evolution for the cavity modes in our problem, the retro-reflection of the external beam splitter adds a complication because the source from the CW mode is then reflected back, coupling to the CCW mode.  This cascading feed-forward results in a modification to the above system's HLE.  Under the limit when the time delay of the retro-reflection is short when compared to our ``slowly-varying'' input fields, the cascading can be approximated as taking place point-wise in time.  Namely, if the optical path length $\Delta L$ from the CW cavity output to the reflecting mirror and the travel time $\Delta t = \Delta L/c$ are both negligibly small, the right-side output is the sum of three distinct terms.  The first is $\hat{a}_r^{in}(t)$ having reflected off the BS. The second is the transmission through the BS of the left-side input, $\hat{a}_l^{in}(t - \Delta L/c) \approx \hat{a}_l^{in}(t)$.  The third is the transmission of the CW source, $L_{cw}(t - \Delta L/C) \approx L_{cw}(t)$.  Thus in total we have the approximate output field, 
\begin{equation}
    \hat{a}_{r}^{out}(t) \approx -\rho\, \hat{a}_r^{in}(t) + \tau e^{i \phi } \hat{a}_{l}(t) + \tau e^{i \phi}\, L_{cw}(t).
\end{equation}

In addition to modifying $\hat{a}^{out}_r(t)$, the retro-reflection modifies both the left-side output as well as the input to the CCW mode.  Thus, if we evaluate the guided field mode to the right of the cavity but still before the BS, we have
\begin{equation}
    \hat{a}_{cw}^{out}(t) = \hat{a}^{in}_l(t) + L_{cw}(t),
\end{equation}
where $L_{cw}(t) = \sqrt{\gamma}\, \hat{a}_{cw}(t)$. 

At this same point, the left-propagating input into the CCW mode is 
\begin{equation}
\begin{split}
\hat{a}_{ccw}^{in}(t) &= \rho e^{2 i \phi} \hat{a}_{cw}^{out}(t - 2 \Delta L/c ) +  \tau e^{i \phi} \hat{a}_r(t - \Delta L/c) \\
\hat{a}_{ccw}^{in}(t) &\approx \rho e^{2 i \phi} \left[ \hat{a}^{in}_l(t) + L_{cw}(t) \right] + \tau e^{i \phi} \hat{a}_r(t),
\end{split}
\end{equation}
The right-propagating input into the CW mode is unaffected by the BS, thus we have $\hat{a}_{cw}^{in}(t) = \hat{a}_l^{in}(t)$. 

Substituting the above expression for $\hat{a}_{cw}^{in}(t)$ and $\hat{a}_{ccw}^{in}(t)$ into the general HLE, Eq.~(\ref{eq:HLeqs}), and then collecting terms proportional to $\hat{\mathbf{a}}^{in}(t) \equiv \left[ \begin{array}{c} \hat{a}^{in}_l(t)\\ \hat{a}^{in}_r(t) \end{array}\right]$, 
results in a HLE of the same form as Eq. (\ref{eq:HLeqs}) but now with effective parameters, i.e., with a new triple $(\mathbf{S}_{eff}, \mathbf{L}_{eff}, H_{eff} )$~\cite{gough_principles_2012, combes_slh_framework_modeling_2017, cook_input-output_2018}, where
\begin{subequations}\label{eq:SLHeff}
\begin{equation}
    \mathbf{S}_{eff} = \mathbf{S} = \left[ \begin{matrix} 
    \rho e^{2 i \phi} & \tau e^{i \phi} \\
    \tau e^{i \phi} & - \rho \end{matrix} \right],
\end{equation}
\begin{equation}\label{eq:Leff}
\begin{split}
    \mathbf{L}_{eff} &= \left[\begin{matrix}\sqrt{\gamma}\, \rho e^{2 i \phi} \hat{a}_{cw} + \sqrt{\gamma}\, \hat{a}_{ccw}\\\sqrt{\gamma} \tau e^{i \phi}\, \hat{a}_{cw}\end{matrix}\right]\\
    &= \mathbf{S B}^{\dag}\cdot \hat{\mathbf{a}}^{cav}\,
\end{split}
\end{equation}
and
\begin{equation} \label{eq:Heff}
\begin{split}
    H_{eff} &= \hat{\mathbf{a}}^{cav\dag} \cdot \left[\begin{matrix}0 & \epsilon + \frac{i \gamma \rho e^{- 2 i \phi}}{2}\\\epsilon - \frac{i \gamma \rho e^{2 i \phi}}{2} & 0\end{matrix}\right] \cdot \hat{\mathbf{a}}^{cav}\\
    &\equiv \hat{\mathbf{a}}^{cav\dag} \cdot \mathbf{H}_{eff} \cdot \hat{\mathbf{a}}^{cav}.
\end{split}
\end{equation}
\end{subequations}
The second line in Eq.~(\ref{eq:Leff}) can be verified from the definitions of $\mathbf{S}$ and $\mathbf{B}$ in Eq.~(\ref{eq:S_matrix}) and ~Eq.~(\ref{eq:B}), respectively.
We note that the complex-valued matrix $\mathbf{H}_{eff}$ is Hermitian.  
The non-Hermitian matrix $\tilde{\mathbf{H}}_\epsilon$ and ``source'' matrix $\mathbf{B}$ can both be derived by evaluating the HLE in Eq.~(\ref{eq:HLeqs}) for $X=\hat{\mathbf{a}}^{cav}$ with parameters given by Eq.~(\ref{eq:SLHeff}).
We note that the input-output relation of Eq.~(\ref{eq:IO_relation_general}) does indeed correspond to Eq.~(\ref{eq:io_t}) for the above 
$\mathbf{L}_{eff}$.  
Furthermore it is also straight forward to compute that for any complex valued $2\times2$ matrix $[m_{jk}]$, we have the commutation relations:
\begin{equation}
\begin{split}
    \left[\hat{a}_i,\, m_{jk}\, \hat{a}_k \right] &= 0 \\
    \left[\hat{a}_i,\, \hat{a}^{\dag}_{j}\, m_{jk} \right] &= m_{ik} \\
    \left[\hat{a}_i,\, \hat{a}_j^{\dag}\, m_{jk}\,\hat{a}_k \right] &= m_{i k}\, \hat{a}_k,
\end{split}    
\end{equation}
where any repeated index has an implied sum.  Using this result we conclude that
\begin{equation}\label{eq:SLH_eff_cr}
\begin{split}
    \left[\hat{\mathbf{a}}^{cav},\, \mathbf{L}_{eff} \right] &= \mathbf{0} \\
    \left[\hat{\mathbf{a}}^{cav},\, \mathbf{L}^{\dag}_{eff} \right] &= \mathbf{B S}^\dag \\
    \left[\hat{\mathbf{a}}^{cav},\, H_{eff} \right] &=-\mathbf{H}_{eff} \cdot \hat{\mathbf{a}}^{cav}.
\end{split}
\end{equation}
where $\mathbf{0}$ is the zero matrix.  Substituting these commutation relations into Eq.~(\ref{eq:HLeqs}) with $\mathbf{L} \mapsto \mathbf{L}_{eff}$ and $X = \hat{\mathbf{a}}^{cav}$, we find that
\begin{equation} \label{eq:a_cav_HLE}
\begin{split} 
    \tfrac{d}{dt} \hat{\mathbf{a}}^{cav} &= -i \mathbf{H}_{eff} \hat{\mathbf{a}}^{cav} - \tfrac{1}{2} \mathbf{B} \mathbf{S}^\dag \mathbf{S}\mathbf{B}^\dag \hat{\mathbf{a}}^{cav} - \mathbf{B} \mathbf{S}^\dag \mathbf{S} \hat{\mathbf{a}}^{cav}\\
    &= -i \left( \mathbf{H}_{eff} - \tfrac{i}{2} \mathbf{B} \mathbf{B}^\dag \right) \hat{\mathbf{a}}^{cav} - \mathbf{B} \hat{\mathbf{a}}^{cav}.
\end{split}
\end{equation}
It is then evident that this is equivalent to Eq.~(\ref{eq:cav_HOM}) if 
\begin{equation}
    \tilde{\mathbf{H}}_{\epsilon} = \mathbf{H}_{eff} - \tfrac{i}{2} \mathbf{B}\mathbf{B}^\dag.
\end{equation}
An exercise in matrix multiplication using Eqs. (\ref{eq:B}) and (\ref{eq:HLeqs}) shows that this is indeed true.

\section{Derivation of $\mathbf{A}$\label{app:A}}

First note that from $\mathbf{K} = \mathbf{S}\left(\mathbf{I} + i \mathbf{B^\dag R B}\right)$, if $\mathbf{K}$ is unitary, we have
\begin{equation}
    \mathbf{K}^\dag\mathbf{K} = \mathbf{I} - i\mathbf{B^\dag R^\dag B} + i\mathbf{B^\dag R B} + \mathbf{B^\dag R^\dag B B^\dag R B} = \mathbf{I}.
\end{equation}
Stripping away the superfluous elements results in the useful relation:
\begin{equation} \label{eq:RdBBdR}
    i \mathbf{R^\dag} - i \mathbf{R} = \mathbf{R^\dag B B^\dag R} =  \mathbf{R B B^\dag R^\dag}. 
\end{equation}
Acting $\mathbf{\tilde{H}}^\dag - \omega \mathbf{I}$ from the left and $\mathbf{\tilde{H}} - \omega \mathbf{I}$ on the right shows that unitarity is guaranteed when 
\begin{equation}
    \operatorname{Im}\,\mathbf{\tilde{H}}_\epsilon = -\tfrac{1}{2} \mathbf{B}\mathbf{B}^\dag.
\end{equation}
In other words, when all of the loss in the cavity is directed to the output modes through $\mathbf{B}^\dag$ with the required input coupling via $\mathbf{B}$, then the overall IO relation is unitary. 

The only distinguishing element between the transfer matrices $\mathbf{K}_{\epsilon +\delta\epsilon}$ and $\mathbf{K}_{\epsilon}$ appears via the resolvent of the effective Hamiltonian $\mathbf{\tilde H}_\epsilon$. 

\begin{widetext}
Computing the difference
\begin{equation}
    \mathbf{\tilde H}_{\epsilon + \delta\epsilon} - \mathbf{\tilde H}_{\epsilon} = \delta\epsilon \mathbf{V}, 
\end{equation}
We find
\begin{equation}
    \begin{split}
    \mathbf{R}_{\epsilon}(z) \left( \mathbf{\tilde H}_{\epsilon + \delta\epsilon} - \mathbf{\tilde H}_{\epsilon} \right) \mathbf{R}_{\epsilon + \delta\epsilon}(z) &=  \delta \epsilon\, \mathbf{R}_{\epsilon}(z) \mathbf{V} \mathbf{R}_{\epsilon + \delta\epsilon}(z) \\
    \mathbf{R}_{\epsilon}(z) - \mathbf{R}_{\epsilon + \delta\epsilon}(z) &= \delta \epsilon\,\mathbf{R}_{\epsilon}(z) \mathbf{V} \mathbf{R}_{\epsilon + \delta\epsilon}(z) \\
    \mathbf{R}_{\epsilon}(z) &= \left( \mathbf{I} + \delta \epsilon\,\mathbf{R}_{\epsilon}(z) \mathbf{V} \right) \mathbf{R}_{\epsilon + \delta\epsilon}(z) \\
\end{split}
\end{equation}
When $\left( \mathbf{I} + \delta \epsilon\,\mathbf{R}_{\epsilon}(z) \mathbf{V} \right) $ is invertible, i.e., the Neumann series $\sum_{n = 0}^\infty\left(-\delta \epsilon \mathbf{R}_{\epsilon}(z) \mathbf{V} \right)^{n}$ converges to $(\mathbf{I} + \delta \epsilon\,\mathbf{R}_{\epsilon}(z) \mathbf{V} )^{-1}$, we then have the relation:
\begin{equation}
\begin{split}
    \mathbf{R}_{\epsilon + \delta\epsilon} &= 
    (\mathbf{I} + \delta \epsilon\,\mathbf{R}_{\epsilon}\mathbf{V} )^{-1} \mathbf{R}_{\epsilon} \\
    &= \mathbf{R}_{\epsilon} - \delta \epsilon\,\mathbf{R}_{\epsilon} \mathbf{V}\mathbf{R}_{\epsilon} + \delta \epsilon^2 (\mathbf{R}_{\epsilon} \mathbf{V})^2 \mathbf{R}_\epsilon + O(\delta \epsilon^3).
\end{split}
\end{equation}
We now want to apply the above expansion to the matrix product $\mathbf{K}^\dag_\epsilon \mathbf{K}_{\epsilon + \delta\epsilon}$.  First we will write 
\begin{equation}
\begin{split}
    \mathbf{K}^\dag_{\epsilon} \mathbf{K}_{\epsilon + \delta\epsilon } &= \left( \mathbf{I} - i \mathbf{B^\dag R^\dag_{\epsilon} B}\right) \left( \mathbf{I} + i \mathbf{B^\dag R_{\epsilon + \delta\epsilon} B}\right) \\
    &=\mathbf{I} - i \mathbf{B^\dag R^\dag_{\epsilon} B} + i \mathbf{B^\dag R_{\epsilon + \delta\epsilon} B} 
    + \mathbf{B^\dag R^\dag_{\epsilon} BB^\dag R_{\epsilon + \delta\epsilon} B}.
\end{split}
\end{equation}
Then expanding $\mathbf{R}_{\epsilon + \delta\epsilon}$ to second order in $\delta\epsilon$ results in
\begin{equation}\label{eq:KdK_eps}
\begin{split}
    \mathbf{K}^\dag_{\epsilon} \mathbf{K}_{\epsilon + \delta\epsilon } &= \mathbf{I} +\mathbf{B}^\dag \left( - i \mathbf{ R}^\dag_{\epsilon}  + i \mathbf{R}_{\epsilon} + \mathbf{ R^\dag_{\epsilon} BB^\dag R_{\epsilon} } \right) \mathbf{B} - \delta\epsilon\, \mathbf{B}^\dag \left( i  \mathbf{R_{\epsilon}}  + \mathbf{R^\dag_\epsilon BB^\dag R_\epsilon } \right) \mathbf{V R_\epsilon B}\\
    &\  + \delta\epsilon^2\, \mathbf{B}^\dag \left( i \mathbf{R_\epsilon } + \mathbf{R^\dag_\epsilon BB^\dag R_\epsilon} \right) (\mathbf{V R_\epsilon})^2 \mathbf{B} + O(\delta\epsilon^3).
\end{split}
\end{equation}
Using Eq.(\ref{eq:RdBBdR}) this simplifies as
\begin{equation}
\begin{split}
    \mathbf{K}^\dag_{\epsilon} \mathbf{K}_{\epsilon + \delta\epsilon } &= \mathbf{I} - i\, \delta\epsilon\, \mathbf{B^\dag R_\epsilon^\dag V R_\epsilon B }\\
    &\quad + i\, \delta\epsilon^2\, \mathbf{B^\dag R^\dag_\epsilon}(\mathbf{V R_\epsilon})^2 \mathbf{B} + O(\delta\epsilon^3).
\end{split}
\end{equation}
Defining the matrix
\begin{equation}
    \mathbf{A} \equiv -\mathbf{B^\dag R_\epsilon^\dag V R_\epsilon B},
\end{equation}
and using the relation
\begin{equation}
    \frac{\partial}{\partial \epsilon} \mathbf{R}_\epsilon = -\mathbf{R_\epsilon V R_\epsilon},
\end{equation}
shows that
\begin{equation}\label{eq:dA_dth}
    \frac{i}{2} \frac{\partial \mathbf{A}}{\partial \epsilon} = \frac{i}{2}\mathbf{B^\dag}\big(\mathbf{ R_\epsilon^\dag V }\big)^2  \mathbf{\mathbf{R}_\epsilon B} + \frac{i}{2}\mathbf{B^\dag R_\epsilon^\dag}\big( \mathbf{V R_\epsilon} \big)^2 \mathbf{B}.
\end{equation}
Manipulating Eq.~(\ref{eq:RdBBdR}) shows
\begin{equation}
\begin{split}
    \frac{i}{2} \mathbf{R_\epsilon^\dag} = \frac{i}{2} \mathbf{R_\epsilon} + \frac{1}{2} \mathbf{R_\epsilon B B^\dag R_\epsilon^\dag},
\end{split}
\end{equation}
which when combined with Eq.~(\ref{eq:dA_dth}) yields
\begin{equation}
    \frac{i}{2} \frac{\partial \mathbf{A}}{\partial \epsilon} = \frac{1}{2} \mathbf{A}^2 + i \mathbf{B^\dag R_\epsilon^\dag}\big( \mathbf{V R_\epsilon} \big)^2 \mathbf{B}. 
\end{equation}
Thus, we have the simplification presented in the main text:
\begin{equation} \label{eq:KdK_expanded}
    \mathbf{K^\dag_\epsilon K_{\epsilon + \delta\epsilon}} = \mathbf{I} + i\, \delta\epsilon \,\mathbf{A} - \left(\mathbf{A}^2  - i \frac{\partial \mathbf{A}}{\partial \epsilon} \right)\frac{\delta\epsilon^2}{2} + O(\delta\epsilon^3).
\end{equation}

A corollary of this is that
\begin{equation} \label{eq:dK/deps}
    \frac{\partial\mathbf{K}_\epsilon}{\partial\epsilon} = +i \mathbf{K}_\epsilon \mathbf{A}_\epsilon
\end{equation}
which follows by acting $\mathbf{K}_\epsilon/\delta \epsilon$ from the left on the result of subtracting $\mathbf{I}$ from Eq.~(\ref{eq:KdK_expanded}), and then taking the $\delta\epsilon \rightarrow 0$ limit. 

\section{NOON state fidelity}\label{app:NOON}
    
Starting from the expansion of $\mathbf{K}_{\epsilon}^\dag \mathbf{K}_{\epsilon + \delta\epsilon}$ in $\delta\epsilon$, the NOON state overlap from Eq.~(\ref{eq:NOON_wick}) is
\begin{equation}
    \langle \Psi_{\epsilon} | \Psi_{\epsilon + \delta\epsilon} \rangle =
    \frac{1}{2} \sum_{n,m = 1}^2 \Big( \langle\boldsymbol{\psi}_n, \boldsymbol{\psi}_m\rangle + i \langle \boldsymbol{\psi}_n, \mathbf{A}_\epsilon \boldsymbol{\psi}_m \rangle \, \delta\epsilon 
    - \langle \boldsymbol{\psi}_n, \mathbf{A}_\epsilon^2 \boldsymbol{\psi}_m \rangle \frac{\delta\epsilon^2}{2} + i\langle \boldsymbol{\psi}_n, \frac{\partial \mathbf{A}_\epsilon}{\partial \epsilon} \boldsymbol{\psi}_m \rangle \frac{\delta\epsilon^2}{2} + \dots \Big)^N.
\end{equation}

Expanding the multinomial up to second order in $\delta\epsilon$ shows that
\begin{equation}
\begin{split}
    \langle \Psi_{\epsilon} | \Psi_{\epsilon + \delta\epsilon} \rangle = 1 + 
    \frac{1}{2} \sum_{n,m = 1}^2\Big(& i N\,\delta\epsilon \, \langle \boldsymbol{\psi}_n, \mathbf{A}_{\epsilon} \boldsymbol{\psi}_m \rangle\, \delta_{ij}^{N-1} - \delta\epsilon^2 \frac{N(N-1)}{2} \langle \boldsymbol{\psi}_n, \mathbf{A}_{\epsilon} \boldsymbol{\psi}_m \rangle^2\, \delta_{ij}^{N-2}\\
    &- N\frac{\delta\epsilon^2}{2}\langle \boldsymbol{\psi}_n, \mathbf{A}_{\epsilon}^2 \boldsymbol{\psi}_m \rangle\, \delta_{ij}^{N-1}  + i N \frac{\delta\epsilon^2}{2} \langle \boldsymbol{\psi}_n, \frac{\partial \mathbf{A}_{\epsilon}}{\partial \epsilon} \boldsymbol{\psi}_m \rangle\, \delta_{ij}^{N-1}  + O(\delta\epsilon^3) \Big).
\end{split}
\end{equation}
Computing the square magnitude,
\begin{equation}
\begin{split}
    |\langle \Psi_{\epsilon} | \Psi_{\epsilon + \delta\epsilon} \rangle|^2 &= 1 +
    \frac{\delta\epsilon^2 N^2}{4} \left|\sum_{n,m = 1}^2  \langle \boldsymbol{\psi}_n, \mathbf{A}_\epsilon \boldsymbol{\psi}_m \rangle\, \delta_{ij}^{N-1}\right|^2\\
    &\quad- \frac{\delta\epsilon^2}{2} \sum_{n,m = 1}^2\Big( N(N-1)\, \langle \boldsymbol{\psi}_n, \mathbf{A}_\epsilon \boldsymbol{\psi}_m \rangle^2\, \delta_{ij}^{N-2} + N \langle \boldsymbol{\psi}_n, \mathbf{A}_\epsilon^2 \boldsymbol{\psi}_m \rangle\, \delta_{ij}^{N-1} \Big) + O(\delta\epsilon^3).
\end{split}
\end{equation}

A useful way of representing this is to consider the N-fold tensor product of the single particle states $\boldsymbol{\psi}_n$, so that we have the $N$ particle state $|\boldsymbol{\psi}_n^{\otimes N} \rangle$, and the collective operators $\mathbf{A}^{tot}_\epsilon = \sum_{n = 1}^N \mathbf{A}^{(n)}_\epsilon$, where $\mathbf{A}_\epsilon^{(n)}$ applies the matrix $\mathbf{A}_\epsilon$ to the $n$th copy and acts as the identity on the remainder.   In this notation, we then have the NOON state fidelity 
\begin{equation}
\begin{split}
    |\langle \Psi_{\epsilon} | \Psi_{\epsilon + \delta\epsilon} \rangle|^2 &= 1 +
    \frac{\delta\epsilon^2}{4} \Big| \sum_{n,m = 1}^2  \langle \boldsymbol{\psi}_n^{\otimes N}, \mathbf{A}^{tot}_\epsilon\, \boldsymbol{\psi}_m^{\otimes N} \rangle \Big|^2 - \frac{\delta\epsilon^2}{2} \sum_{n,m = 1}^2  \langle \boldsymbol{\psi}_n^{\otimes N}, \big( \mathbf{A}^{tot}_\epsilon\big)^2\,  \boldsymbol{\psi}_m^{\otimes N} \rangle + O(\delta\epsilon^3),\\
    & = 1 - \delta\epsilon^2 \left(\Delta \mathbf{A}^{tot}_\epsilon \right)^2 + O(\delta\epsilon^3),
\end{split}
\end{equation}
where the variance $(\Delta \mathbf{A}^{tot}_\epsilon )^2$ is computed under the $N$ particle superposition: $|NOON\rangle = \frac{1}{\sqrt{2}} \left( \psi_1^{\otimes N} + \psi_2^{\otimes N}\right) $.

\end{widetext}

\section{Relation of the QFI to the SNR in Ref.~\onlinecite{lau_fundamental_2018}}\label{app:Lau_clerk}

The fundamental assumption of Lau and Clerk's analysis is that under a small perturbation $\epsilon$, the reflected output state can be written as the expectation value at $\epsilon = 0$, plus a linear shift proportional to $\epsilon$.  In other words, they assume that the reflected output field operator $\hat{a}^{out}(t)$  has an expectation value that is a linear shift, 
\begin{equation}\label{eq:lc_shift}
    \left\langle a^{out}(t)\right\rangle_\epsilon = \left\langle a^{out}(t)\right\rangle_0 + \lambda\,\epsilon,
\end{equation}
for some complex coefficient $\lambda$.  The instantaneous homodyne current is modeled as measuring the operator
\begin{equation}
    \hat{I}^{out}(t) \equiv \sqrt{\frac{\gamma}{2} }\left( e^{-i\varphi} \hat{a}^{out}(t) + e^{i\varphi} \hat{a}^{out\,\dag}(t)\right),
\end{equation}
for a quadrature phase $\varphi$ that will optimize the SNR. The signal in the SNR, $\mathcal{S}$, is defined to be the squared deviation in the integrated homodyne current induced by $\epsilon$, i.e., 
\begin{equation}
    \mathcal{S} \equiv \left(\int_0^\tau dt\, \langle I^{out}(t) \rangle_\epsilon - \langle I^{out}(t) \rangle_0 \right)^2,
\end{equation}
where $\tau$ is the total integration time.  Substituting $\hat{I}^{out}(t)$ and the linear shift assumption results in
\begin{equation}
    \mathcal{S} = \frac{\gamma}{2}\left[\int_0^\tau dt\, 2 \epsilon\, \operatorname{Re} (e^{-i \varphi} \lambda ) \right]^2 \le 2 \gamma \epsilon^2\, |\lambda|^2\, \tau^2.
\end{equation}
Thus the optimal phase angle $\varphi$ which saturates the inequality exactly matches the phase $\lambda$, so that $e^{i\varphi} = \lambda/|\lambda|$.  

The noise factor $\mathcal{N}$ is limited by the fundamental delta-correlated white noise associated with vacuum input noise.  In terms of the homodyne measurement, the instantaneous current will have a delta correlated input of $\left \langle \hat{I}^{out}(t) \hat{I}^{out}(t') \right \rangle_{\epsilon} = \frac{\gamma}{2} \delta( t- t')$.  When integrated over the duration $\tau$, the signal $\mathcal{S}$ has the associated noise $\mathcal{N} = \gamma \tau /2$, so that the minimum SNR is
\begin{equation}
    \mathcal{S}/\mathcal{N} = 4 \tau\, \epsilon^2\, |\lambda|^2.
\end{equation}
Of the three parameters in the SNR, only $\lambda$ remains unspecified, and thus a limiting bound on the SNR is equivalent to a bound on $|\lambda|$.  

In order to identify a relationship between $\lambda$ and the QFI, we must first cast the integrated homodyne current into the frequency domain.  For a probe coherent state amplitude which is nearly all contained within the interval $[0, \tau]$ we have
\begin{equation}
\begin{split}
    \int_0^\tau dt\, \langle \hat{I}^{out} (t)\rangle_{\epsilon} &= \sqrt{\frac{\gamma}{2}} \int_0^\tau dt\, e^{-i\varphi} \langle \hat{a}^{out}(t) \rangle_\epsilon + c.c. \\
    &= \sqrt{\frac{\gamma}{2}}\,e^{-i\varphi}\, \int dt\, \beta^{out}_\epsilon (t)  + c.c.\\
    &= \sqrt{\frac{\gamma}{2}} e^{-i\varphi} \beta^{out}_{\epsilon}(\omega)|_{\omega = 0} + c.c. \\
    &= \sqrt{\frac{\gamma}{2}} e^{-i\varphi} [ \mathbf{K}_\epsilon \boldsymbol{\beta}^{in}]_{1}(0) + c.c.,
\end{split}    
\end{equation}
where $\boldsymbol{\beta}^{in}(\omega)$ is the frequency domain vector of input amplitudes.  

The assumption of a linear shift is equivalent to expanding $\mathbf{K}_{\epsilon}$ to only first order in $\epsilon$, which from Eq.~(\ref{eq:dK/deps}) results in $\mathbf{K}_{\epsilon} = \mathbf{K}_0 + i \mathbf{K}_0\mathbf{A}_0\,\epsilon  + O(\epsilon^2)$.  Therefore
\begin{equation}
    \mathcal{S} = \frac{\gamma \epsilon^2}{2}\Big(  i e^{-i\varphi} [ \mathbf{K}_0 \mathbf{A}_0 \boldsymbol{\beta}^{in}]_{1}(0) + c.c. \Big)^2.
\end{equation}
If we again set $\boldsymbol{\beta}^{in}$ to correspond to the eigenstate of $\mathbf{A}_\epsilon$ with eigenvalue $\lambda_{abs}$, $\boldsymbol{\beta}_{abs}$,
\begin{equation}
\begin{split}
    \mathcal{S} &= \frac{\gamma \lambda_{abs}^2 \epsilon^2}{2}\Big(  i e^{-i\varphi} [ \mathbf{K}_0 \boldsymbol{\beta}_{abs}]_{1}(0) + c.c. \Big)^2 \\
    &\le 2\, \gamma \lambda_{abs}^2 \epsilon^2\, |\beta_{abs}(0)|^2,
\end{split}
\end{equation}
where the inequality comes from matching the homodyne phase $\varphi$ to equal the phase $\operatorname{arg}\, i [\mathbf{K}_0 \boldsymbol{\beta}^{in}]_{1}(0)$. 

If in the time domain the input coherent state is a resonant square pulse of duration $\tau$, and average photon flux $|\beta_{lc}|^2$, then moving to the frequency domain results in 
\begin{equation}
    |\beta_{abs}(0)|^2 = |\beta_{lc}|^2\, \tau^2 = \|\boldsymbol{\beta}_{abs} \|^2\, \tau.
\end{equation}
Combining all these elements to compute the SNR, we find
\begin{equation}
    \begin{split}
    \mathcal{S}/\mathcal{N} &= \frac{2 \gamma \lambda_{abs}^2\, \epsilon^2 \|\boldsymbol{\beta}_{abs} \|^2\, \tau}{\gamma \tau /2} \\
    &= 4\, \lambda_{abs}^2\, \|\boldsymbol{\beta}_{abs}\|^2\, \epsilon^2.
    \end{split}
\end{equation}
Thus we conclude that the SNR  of Ref.~\onlinecite{lau_fundamental_2018} is simply equal to the coherent state QFI, $\mathcal{I}_\beta(0)$, multiplied by $\epsilon^2$. 


%

%

\end{document}